\pdfoutput=1
\documentclass[10pt,a4paper]{article}      % Specifies the document class

\usepackage{amsthm,amsmath,amssymb,xcolor,sectsty,latexsym,mathtools,mathrsfs}
\usepackage{slashed}  
\usepackage{bm}
\usepackage{epsfig}
\usepackage{dcolumn}
\definecolor{vermelho}{cmyk}{0,.88,.77,.40}
\usepackage{cite}
\usepackage{color}
\numberwithin{equation}{section}
\usepackage[textwidth=6.3in,top=1.3in,bottom=1.3in]{geometry}
\usepackage{setspace}
\chapterfont{\color{vermelho}}

\newcommand{\be}{\begin{equation}}

\newcommand{\ee}{\end{equation}}
\newcommand{\beq}{\begin{equation}}
\newcommand{\eeq}{\end{equation}}  
\newcommand{\ba}{\begin{eqnarray}}
\newcommand{\ea}{\end{eqnarray}}
\newcommand{\bef}{\begin{figure}}
\newcommand{\eef}{\end{figure}}

\newcommand{\p}{\partial}

\newcommand{\si}{\sigma}

\newcommand{\g}{\gamma}

\newcommand{\nn}{\nonumber}

\begin{document}

\nopagebreak

\title{  \begin{center}\bf Thermalized Axion Inflation \end{center} }

\vfill
\author{Ricardo Z. Ferreira \footnote{rferreira@icc.ub.edu}, ~ Alessio Notari \footnote{notari@ub.edu}
}
\date{ }

\maketitle

\begin{center}
	\vspace{-0.7cm}
	{\it  Departament de F\'isica Qu\`antica i Astrof\'isica i Institut de Ci\`encies del Cosmos (ICCUB)}\\
	{\it  Universitat de Barcelona, Mart\'i i Franqu\`es, 1, E-08028, Barcelona, Spain}\\
	\vspace{0.2cm}
	
\end{center}

\begin{abstract}
We analyze the dynamics of inflationary models with a coupling of the inflaton $\phi$ to gauge fields of the form $\phi F \tilde{F}/f$, as in the case of axions. It is known that this leads to an instability, with exponential amplification of gauge fields, controlled by the parameter $\xi= \dot{\phi}/(2fH)$, which can strongly affect the generation of cosmological perturbations and even the background. We show that scattering rates involving gauge fields can become larger than the expansion rate $H$, due to the very large occupation numbers, and create a thermal bath of particles of temperature $T$ during inflation. In the thermal regime, energy is transferred to smaller scales, radically modifying the predictions of this scenario. We thus argue that previous constraints on $\xi$ are alleviated.  If the gauge fields have Standard Model interactions, which naturally provides reheating, they thermalize already at $\xi\gtrsim2.9$, before perturbativity constraints and also before backreaction takes place. In absence of SM interactions ({\it i.e.} for a dark photon), we find that gauge fields and inflaton perturbations thermalize if $\xi\gtrsim3.4$; however, observations require $\xi\gtrsim6$, which is above the perturbativity and backreaction bounds and so a dedicated study is required. After thermalization, though, the system should evolve non-trivially due to the competition between the instability and the gauge field thermal mass. If the thermal mass and the instabilities equilibrate, we expect an equilibrium temperature of $T_{eq} \simeq \xi H/\bar{g}$ where $\bar{g}$ is the effective gauge coupling. Finally, we estimate the spectrum of perturbations if $\phi$ is thermal and find that the tensor to scalar ratio is suppressed by $H/(2T)$, if tensors do not thermalize.

\end{abstract}

\section{Introduction}
 
Inflation is a successful paradigm for the Early Universe, which provides consistent initial conditions for the radiation era and a spectrum of cosmological perturbations in agreement with observations. Its most common realization is thought as a cold state, with a very long stage of quasi exponential expansion due to a scalar field slowly rolling down a flat potential. Such a dynamics exponentially dilutes any remnant but should still be able to reheat the universe at the end of inflation and provide the hot radiation era through some couplings of the scalar field to the Standard Model fields.
 
A plethora of slow-roll models has been studied within this paradigm, with more or less agreement with observational data. 
Here we wish to explicitly show the viability of a physically more rich possibility, namely that of a hot plasma already present during inflation, therefore unifying inflation and reheating, and opening thus a new class of inflationary models with its own peculiar predictions. One main purpose of this paper is to construct a working model in which the field can indeed dynamically create such a plasma. 

Such a possibility has also been considered by~\cite{Berera1995,Berera2008} under the name of {\it warm inflation}, by invoking a dissipation term due to a coupling to a thermal bath of particles\footnote{See also \cite{Morikawa:1984dz, Sakagami:1984ae} for related earlier work.}. An obvious difficulty is the need of an exponential production of radiation in order to overcome the exponential dilution without spoiling the slow-roll stage. To achieve this goal we propose a {\it thermalized axion inflation} model in which we simply couple gauge-fields $A_\mu$ to an axion-like field $\phi$ (which we will think of as the inflaton), through an axial coupling. 
The phenomenology of such a coupling during inflation has been frequently studied in the literature (see \cite{Anber2009, Barnaby2011, Linde2012, Ferreira2014a, Ferreira2015a, Peloso2016, Notari2016} for an incomplete list of references). 

The interest about this coupling lies in the instability it triggers on the gauge fields equation of motion in presence of a constant field velocity $\dot{\phi}$, leading to strong particle production, that starts at wavelengths of ${\cal O}((\xi H)^{-1})$, slightly smaller the horizon size. This instability is present already at linear order in $\dot{\phi}$; the deep reason behind this fact is that the Lagrangian couples $\phi$ to a CP-odd (and thus T-odd) term. 
Another interesting feature of such a coupling is that the gauge field production can become so large that it backreacts on the background and dynamically generates slow-roll even in absence of a flat potential~\cite{Anber2009,Notari2016}. This happens when the parameter that controls the instability, $\xi\equiv \dot{\phi}/(2 f H)$, is large enough. It is unclear how to reliably compute the behavior of perturbations in such a backreacting regime, since the gauge field can also backreact on perturbations. In absence of backreaction there are bounds on $\xi$: it has been shown that at large $\xi$ the gauge fields can leave a large non-Gaussian effect in the curvature perturbation of cosmic microwave background \cite{Barnaby2011, Bartolo2014, Ferreira2014a}. Another important constraint comes from requiring perturbativity in the loop expansion~\cite{Ferreira2015a}. Other constraints were derived from the overproduction of black holes by the same mechanism, assuming a given evolution for $\xi$ as a function of time and assuming to know the behavior of perturbations in the backreacting regime~\cite{Lin:2012gs, Linde2012, Bugaev:2013fya}. In principle one could also think of generating extra tensor modes~\cite{Sorbo2011} through this mechanism, but this becomes difficult due to both non-Gaussianity constraints and the requirement of perturbativity~\cite{Ferreira2015a, Peloso2016}.

The main idea of the present paper is that since the instability is able to produce an enormous amount of gauge fields during inflation, the cross sections for gauge field scatterings are largely enhanced by the occupation numbers and their rates are able to overcome the exponential dilution, thus, naturally leading to thermalization and formation of a hot plasma. The fact that all the dynamics is generated by axion-like particles and gauge fields, both having strong protecting symmetries, makes the whole setup well protected. In particular, the axial coupling respects the shift symmetry of $\phi$ and so all induced quantum and thermal corrections should involve derivatives of $\phi$ and so cannot affect the axion potential even when the field thermalizes\footnote{In the U(1) case this is exact, while in the non-abelian case non-perturbative effects can break this symmetry and only leave a discrete shift symmetry.}. This fact is of great importance as it means that this setup does not introduces new $\eta$-problems\footnote{Note that this the scenario is still sensitive to possible corrections coming from irrelevant operators, as much as, for example, the standard natural inflation setup \cite{Freese:1990rb}. Such operators, if present, could become less problematic in a strongly backreacting case since inflation could happen on a steeper potential, due to a friction effect \cite{Berera:2004vm, Anber2009}. However, the study of such a regime goes beyond the scope of this paper.}.

When thermalization is reached, energy moves from the horizon to smaller scales thus completely changing the predictions of this scenario.  In fact, we will show that at large $\xi$ the inflaton perturbations become thermal inside the horizon, therefore changing the standard vacuum prediction at horizon crossing and, as a consequence, the predictions for cosmological observables. However, as we will see, this happens when backreaction and higher loop corrections become important and so a dedicated study is needed.

Nevertheless, gauge field thermalization can also happen in a perturbative and non-backreacting regime. That is the case if one considers a, probably more realistic, scenario where the gauge fields belong to the Standard Model (SM). In this case the system can easily reach a partially thermalized state, at lower $\xi$, in which gauge fields are thermal, while the inflaton perturbations become thermal only at higher $\xi$.
In this case reheating of the universe is completely unified with inflation and no extra ingredient is needed: radiation era simply starts when the potential driving inflation becomes subdominant compared to radiation energy density~\cite{Cheng:2015oqa, Adshead2015,Notari2016}.
Thus, an interesting question to ask is whether the phenomenological constraints on $\xi$ due to non-Gaussianity can be alleviated. In the thermal regime, because of this transfer of energy to smaller scales, we have good reasons to think that a phenomenologically viable window can exist, although in the present paper we will only give qualitative arguments for this to be the case and postpone a full analysis to future work.

But this is not the end of the story. Even though thermalization can be obtained from the initially large occupation numbers the subsequent evolution can be very non-trivial as a result of the competition between the instability, which also becomes less efficient after thermalization, and the generation of thermal masses for the gauge field. Although this subsequent dynamics requires a dedicated study including all these important effects, we argue why the instability and the presence of thermal masses should balance each other and either lead to some periodic behavior or to some stationary stage at an equilibrium temperature $T_{eq}>H$. Interestingly, in the latter case we derive very interesting predictions for the spectrum of scalar and tensor perturbations.

The structure of the paper is as follows: in section~\ref{model} we summarize the features of the model; in section~\ref{thermalization} we study the onset of thermalization, by writing Boltzmann-like equations for scatterings and decays, both with the axial coupling and the SM couplings; in section~\ref{numerics} we solve numerically the set of Boltzmann equations to verify our expectations; in section~\ref{phenomenology} we study the phenomenological constraints and predictions of the thermalized system; in section~\ref{discussion} we discuss how the presence of thermal masses affects the subsequent evolution of the system; finally in sec.~\ref{conclusions} we draw our conclusions. In appendix \ref{appendix} we present some additional material related to the numerical solutions as well as some additional derivations.

\begin{figure}
	\centering
	\includegraphics[scale=0.4, angle=270]{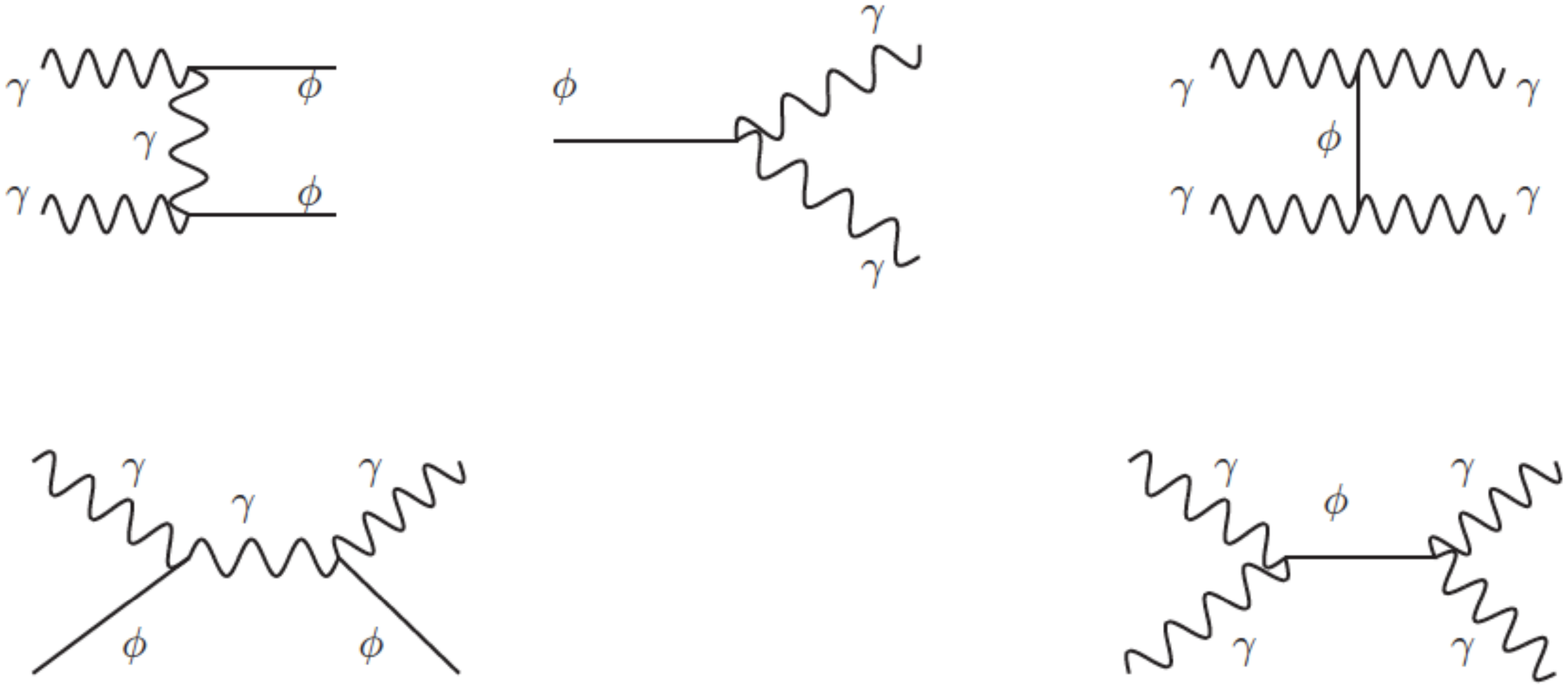}
	\caption{Scatterings and decays induced by the axial coupling and involving gauge fields, $\g$, and the inflaton, $\phi$. \label{diagrams}}
\end{figure}

\section{The axial coupling} \label{model}

Let us consider an axion-like scalar field $\phi$ with a potential $V(\phi)$, coupled to a gauge field $A_\mu$, with field strength  $F_{\mu \nu}$, via an axial coupling, as described by the action
\begin{equation}
	S = \int d^4x \sqrt{-g} \left[\frac{M_p^2}{2} R + \frac 1 2 \p_\mu\phi \p^\mu \phi - V(\phi) -\frac{1}{4} F_{\mu \nu} F^{\mu \nu} - \frac{\phi}{4 f} F_{\mu \nu} \tilde{F}^{\mu \nu} \right].
\end{equation}
Here  $\tilde{F}^{\mu \nu}=  \epsilon^{\mu \nu \alpha \beta} F_{\alpha \beta}  /(2 \sqrt{-g})$ is the dual of $F_{\mu \nu}$. Note that the above coupling induces a periodic potential, of period $f$, only in the non-abelian case, while in the abelian case $V(\phi)$ can be completely unrelated to $f$. Actually, in the context of this paper the only relevant ingredient is the axial coupling to gauge fields, and so our considerations apply also to the case of a non-periodic potential.

We split the scalar field into a spatially homogeneous background value, which drives inflation, and a perturbation, $\phi=\phi (\tau) +\delta\phi(\vec{x},\tau)$ and we approximate the metric with a de-Sitter form in conformal coordinates $ds^2=a^2(\tau)[d\tau^2-d\vec{x}^2]$, where $a=(-H\tau)^{-1}$ and $\tau$ goes from $-\infty$ (past) to $0^-$ (future). Here $H\equiv a'/a^2$ is the Hubble constant during inflation and a prime denotes the derivative taken with respect to conformal time $\tau$. In such a background, the gauge field of comoving momentum $k$ satisfies the following equation of motion in Coulomb gauge (see~\cite{Tkachev:1986tr} and references therein):
\begin{eqnarray} \label{EOM for A}
A_\pm'' + \left(k^2 \pm \frac{2k \xi}{\tau} \right)A_\pm =0,  \qquad \xi\equiv \frac{\dot{\phi}}{ 2 f H} \, ,
\end{eqnarray}
where $\pm$ denotes the positive and negative helicity.
It is immediate to see that the axial coupling triggers an instability at low $k$ in one of the gauge field polarizations, $\gamma_+$, controlled by the dimensionless parameter $\xi$. Instead, the other polarization, $\gamma_-$, gets a different dispersion relation although still positive.

We consider the simple case $\xi \simeq {\rm constant} $, which is a good approximation if $\phi$ is slowly-rolling down its potential. It is also a good approximation if we are in the regime of strong backreaction of the gauge fields on $\phi$~\cite{Anber2009,Notari2016}. In both cases $\xi = \sqrt{\epsilon/2} M_p/f$, where $\epsilon\equiv \dot{\phi}^2 /(2 M^2_p H^2)$ is the first slow-roll parameter. The equation of motion then has analytical solutions that can be written for example by a Whittaker function:
\begin{eqnarray} \label{sol for A}
A_+(k, \tau) = \frac{1}{\sqrt{2k}} e^{\pi\xi/2} W_{-i\xi, 1/2}(2ik \tau) \, ,
\end{eqnarray}
which, in rough terms, connects the flat space oscillatory regime deep inside the horizon\footnote{Note however that even at  $\tau\rightarrow -\infty$ the mode functions always have a logarithmic time-dependent phase.} with an exponential growth for $-k\tau <2\xi$, until the solution approaches a constant $A_k = e^{\pi \xi} /(2 \sqrt{\pi k \xi})$ well outside the horizon, when $-k\tau \lesssim (8\xi)^{-1}$.

\section{Thermalized Axion Inflation} \label{thermalization}

The instability created by the axial coupling leads to a strong production of  gauge fields $\gamma_+$, that starts already at momenta larger than $H$. The goal of this section is to point out that the scatterings and decays involving the gauge field are strongly enhanced by the large particle number and can, in some cases, change dramatically the predictions studied so far in the literature. We will analyze two cases: (1) the gauge field is {\it only} coupled to $\phi$, {\it i.e.} the gauge field is a ``dark photon", not part of the SM; (2) the gauge field belongs to the SM (either abelian or non-abelian) with known interactions to charged particles. The latter case is probably more realistic, because to have successful reheating we would anyway need to couple the system to the SM. 

Crucially the scattering probabilities are Bose-enhanced if the number of $\gamma_+$ is large, which will happen when $\xi \gtrsim 1$. As a consequence, if the particle number overcomes some threshold, which depends on $\xi$ and $f$, the scatterings will lead to an equilibrium distribution and as a result the modes will be redistributed from $k\lesssim 2 \xi H$  to a new scale given by the temperature $T$. Roughly speaking, when the rates $\Gamma_s$ associated to scatterings are larger than $H$ we expect the fields to be in a thermal distribution. 
Such a thermalization involves both polarizations of the gauge field, $\gamma_+$ and $\gamma_-$, as well as $\phi$ or other charged particles (depending on the size of the cross sections) and would lead, {\it a priori}, to a Bose-Einstein (BE) distribution defined as
\begin{eqnarray}
N_\text{BE}(k)= \left(e^{-\frac{\omega(k)}{a T}}-1 \right)^{-1} \, ,
\end{eqnarray}
where $\omega(k)= \sqrt{k^2 + a^2 m^2}$ is the comoving energy of the particle.
Note, however, that $\gamma_+$, due to the imaginary dispersion relation modes and the associated sourcing of low momenta modes $k/a<2 \xi H$, would be infinitely populated if one could wait an arbitrarily long time. In our inflationary background each mode spends a finite amount of time in the instability region and so it never reaches an infinite occupation number but, as a remnant of such dynamics, we expect relevant deviations from a BE distribution, such as the presence of a peak at low momentum. On the other hand, $\gamma_-$ should still be described by a BE distribution with a modified dispersion relation with $\omega^2=k^2+m^2$, where $m^2=-2 k \xi/\tau$ while the $\phi$ fluctuations, if thermalized, are instead described by a massless BE distribution. 

In absence of collisions the energy density in the gauge fields is given by $\rho_R \approx  10^{-4} H^4 e^{2\pi \xi} /\xi^{3}$ \cite{Anber2006}, due to the continuous excitation of modes, as described by eq.~(\ref{EOM for A}). A simple estimate of the expected temperature at thermalization is roughly given by ${\bar T}\approx \rho_R^{1/4}\approx 0.1 H e^{\pi \xi/2}$. Note  that in our system three scales of interest are present: the horizon scale $H$, the instability scale $2 \xi H$ (each mode starts getting excited when its momentum satisfies $k/a<2 \xi H$) and the temperature at thermalization ${\bar T}$. If $\xi\gtrsim {\cal O}(1)$ there is a hierarchy of scales: $H< 2 \xi H< \bar{T}$. We will consider $\xi$ in the range 1$\sim$10: in fact, if it gets very large backreaction will start and one should consider the dynamics of the background together with the mode evolution~\cite{Anber2009,Notari2016}, which we postpone to future work. In any case, even in presence of backreaction $\xi$ should be related logarithmically to the parameters of the potential and typically be at most ${\cal O}(10)$~\cite{Anber2009}.

After thermalization is reached, however, the system can evolve in a non-trivial way. In fact, having an interacting plasma typically implies that gauge fields have thermal masses, which should screen the instability and, as a result, quench the particle production, so that the temperature is expected to decrease. Moreover, the energy extraction from the scalar field is proportional to $\dot{\phi}/(4f) \langle  F \tilde{F} \rangle=a^{-4} \int k d^3 d/d\tau(|A_+|^2-|A_-|^2)$ \cite{Anber2009,Notari2016}; such a quantity should be suppressed if $\gamma_+$ and $\gamma_-$ are in equilibrium, since they tend to compensate each other when averaged over a thermal distribution. One could imagine that the system could reach a stationary configuration at a temperature smaller than $\bar{T}$, or perhaps an oscillatory behavior. We postpone the study of the full evolution after thermalization to future work, although we will comment on some expected features in section~\ref{discussion}.

\subsection{Boltzmann equation and particle numbers} \label{sect:Boltzmann}

In order to study the dynamics that we have just described we define an effective $k$-dependent particle number $N_X(k)$ for each field $X$ and derive the associated Boltzmann-like equations. Written in this form we can then insert the standard flat space scattering terms, multiplied by the appropriate scale factors\footnote{This should be a good approximation as long as the scattering rates are larger than the expansion rate $H$, which is precisely the regime we are interested in.}.

We will only consider subhorizon modes because we expect superhorizon modes to become frozen and not to participate in the collisions. This simplified treatment should give an accurate order of magnitude estimate for the parameters $f$ and $\xi$ such that thermalization happens, while we postpone a more rigorous treatment for future work.

For a given field $X$, with mode functions $X_k$, we define a comoving particle number as the ratio between energy density $\rho_k$ and energy per particle $\omega(k)$:
\begin{eqnarray}
	1/2+N_X(k) \equiv \frac{k^2 |X_k|^2 + |X'_k|^{2}}{2 \, \omega(k)} \, . \label{number}
\end{eqnarray}
Deep inside the horizon, in the Bunch-Davies vacuum, the fields have a clear particle interpretation, since they behave as $X_k=e^{i k \tau}/\sqrt{2 k}$ at $\tau \rightarrow -\infty$, with vacuum particle number $N_{X}=0$. In the scalar field case the previous definition will be valid for the canonically normalized field $u \equiv a \delta \phi$, which satisfies
\begin{eqnarray}
	u_k'' + \left(k^2 - \frac{2}{\tau} \right)u_k =0 \, , \label{ueq}
\end{eqnarray} 
where we neglected, for simplicity, slow-roll corrections. In appendix \ref{bog} we also show that the definition of $N_X(k)$ matches the more standard one given in terms of the Bogolyubov coefficient $N_k =|\beta_k|^2$.

As we mentioned in the beginning of this section, the frequency of a mode inside the instability band is non-trivial and so we need to make an extra assumption in the above definition eq.~(\ref{number}), when specifying the form of $\omega(k)$. In fact, the frequency that appears in the equations of motion is in principle time-dependent and can be imaginary: $\omega_{+,-}^2=(k^2\pm2 k \xi/ \tau)$ for the gauge fields and $\omega_u^2=(k^2-2/\tau^2)$ for the scalar. While in the scalar case $\omega_u^2>0$ inside the horizon and so the notion of particle is well defined, that is not the case for $\omega_{+}$. To overcome this problem, while keeping the treatment simple, we assume in the above definition that $\omega_{+,-}\simeq k$. For the $(-)$ polarization this is a good approximation, especially deep inside the horizon, since as an order of magnitude $\omega_-(k)\simeq k$. However, as we discussed before, for $\gamma_+$ that might not be a good approximation when $-k \tau \simeq 2\xi $, since $\omega_+(k) \rightarrow 0$. Although each comoving mode $k$ is redshifted and so only stays a short amount of time in such regions, to circumvent this problem one can imagine that at each point in time, when the possible thermalization of the system is to be analyzed, $\xi$ is instantaneously driven to zero, either by slowing down $\phi$ or by increasing $f$. In that case the particle number becomes well defined for all modes $k$. In fact in our numerical simulation we obtained very similar results by using this approach: inserting a large particle number for the gauge fields, by using as an initial condition the solution of the equation of motion with a source but without collisions, and then evolving the system for short time scales in absence of the source term.

Using the above definition of effective particle number we are able to rewrite the equations of motion, eq.~(\ref{ueq}) and  eq.~(\ref{EOM for A}), as an equation for the number of right-handed gauge fields, $N_{\g_+}(k)$, and $u$ particles, $N_u(k)$:
\begin{eqnarray}
	N_{\g_+}'(k)  &=& -\frac{4 k \xi}{\tau}  \frac{ \text{Re}\left[ g_A (k, \tau)\right] }{  |g_A (k, \tau)|^2 + k^2} \, \left(N_{\g_+} (k)+1/2\right) \label{eqsnumbers} \, , \\
	N'_{u_{\phantom +}}(k) &=& \frac{4}{\tau^2}  \frac{ \text{Re}\left[ g_u (k, \tau)\right]}{ |g_u (k, \tau)|^2 +k^2} \, \left(N_u (k)+1/2\right) \label{eqsnumbers2} \, , \\
	N_{\g_-}'(k)  &= &0  \label{eqsnumbers1} \, ,  
	\end{eqnarray}
where $g_u \equiv  u' /u$ and $g_A \equiv  A_+' / A_+$. We have also included the left-handed gauge fields, $N_{\g_-}(k)$, which are not sourced and so conserved in the absence of collisions.  
This set of first order Boltzmann-like differential equations is exact and it has a suitable form to include collision terms. Note, however, that in the presence of collisions we would also need to know what happens to the evolution of $g_A$ and $g_u$.  We assume $g_{u,A}$ to be the ones given by the free solution in the absence of scatterings\footnote{Note that in the case of free solutions $g(k, \tau) = - i k$ and so $\text{Re}[g]= 0$. Therefore, what makes the source to be non-zero is precisely the modified dispersion relation. In the case of $u$ this effect is negligible inside the horizon and so we could even have neglected the source for the processes we are interested in.}. This approximation is well justified if we want to capture the onset of thermalization. For $g_u$ we use the well-known exact positive frequency solution of eq.~(\ref{ueq})
\begin{eqnarray}
	u = \frac{e^{i k \tau}}{\sqrt{2 k}} \left (1+\frac{i}{k \tau} \right) \, , \qquad g_u= \frac{i \left(k^2 \tau ^2+i k \tau -1\right)}{\tau  (k \tau +i)} \, , \label{solu}
\end{eqnarray} 
while for $g_A$ we can use the exact solution eq.~(\ref{sol for A}).

\subsection{Collision terms}

Thermalization is triggered by the instability in the gauge fields, which populates the phase space and enhances the collision rates, represented by scatterings and decays. Instead, the instability in the scalars, eq.~(\ref{ueq}), does not play a big role in the thermalization but it is crucial to freeze the perturbations around horizon crossing. 
We insert, thus, collision terms to the right hand side of eqs.~(\ref{eqsnumbers}), (\ref{eqsnumbers1}) and~(\ref{eqsnumbers2}) including scatterings $S_n$ and decays $D$, as:
\begin{eqnarray}
	N_{\g_+}'(k)  &=& -\frac{4 k \xi}{\tau}  \frac{ \text{Re}\left[ g_A (k, \tau)\right] }{  |g_A (k, \tau)|^2 + k^2} \, \left(N_{\g_+} (k)+1/2\right) + S^{++} + S^{+\phi} + D^{+\phi} + S^{+-}  \, ,\label{BoltzA} \\  	N'_{u_{\phantom +}}(k) &=& \frac{4}{\tau^2}  \frac{ \text{Re}\left[ g_u (k, \tau)\right]}{ |g_u (k, \tau)|^2 +k^2} \, \left(N_u (k)+1/2\right) -  S^{+\phi}  - D^{+\phi}  \, ,\label{Boltzu} \\
	N_{\g_-}'(k)  &= &- S^{+-}  \, ,\label{BoltzA2} 
\end{eqnarray}
where the superscript in the collision terms denotes which particles are involved in the process. For completeness we should also include processes involving $\gamma_-$ and $\phi$. However, since $\g_-$ are not produced by the source, such processes should be not relevant to reach thermalization and so we neglect them.

We summarize here all the assumptions used in deriving the above Boltzmann equations:
\begin{itemize}
	\item 
	the function $g_A$ in eq.~(\ref{BoltzA}) is given by the solution in absence of scatterings, which is accurate at least before the onset of thermalization;
	\item 
	the particle number entering in the collision terms is given by eq.~(\ref{number}) with $\omega(k) \simeq k$, as discussed above;
		\item 
	backreaction effects of gauge fields on $\phi$ are neglected for simplicity.
\end{itemize}

\subsubsection*{Scatterings}

In the scatterings we use the flat space cross sections for the canonically normalized fields and use the massless free propagator for both the gauge fields and for $u$. 
Note that, when using the canonical field $u$ the axial vertex gets rescaled by $1/a(\tau)$. We do not treat exactly such factors in the diagrams but we simply multiply each scattering operator by an overall $1/a(\tau)^4$ and assume no other change in the computation. If thermalization happens relatively fast compared to the expansion rate this approximation should be accurate. We, anyway, never run our numerical codes for more than ${\cal O}(1)$ e-fold of expansion.
Since the typical energy is $\omega(k) \lesssim \xi H$, we always assume $\xi H \ll f$, so that we are below the cutoff of the theory. For the same reason one could also require $T \ll f$ because after thermalization modes of energy $\omega(k)  \lesssim T$ are populated; however if this requirement is violated it simply means that one is also exciting modes of other fields, belonging to a UV completion of our effective model (involving for example heavy fermions).

We consider the scatterings shown in fig.~\ref{diagrams}. Each scattering term $S_k$ has the form
\begin{eqnarray}
	S_k = \frac{1}{\omega(k)} \int  \prod_{i=2}^4 \left( \frac{d^3 \vec{k}_i}{(2\pi)^3 (2\omega_i)}  \right) \left| M_i\right| ^2 (2\pi)^4 \delta^{(4)} \left(k^\mu+k_2^\mu-k_3^\mu-k_4^\mu  \right)  B(k,k_2,k_3,k_4) \, ,
\end{eqnarray}
where $k_i^\mu=(\omega_i,\vec{k}_i)$ are the momenta of the external legs, $k_i=|\vec{k_i}|$, $k^\mu=(\omega,\vec{k})$,  $M_n$ is the matrix element of the process (given in appendix \ref{MatrixElements}) and $B$ are the phase space factors  
\begin{eqnarray}
	B (k_1,k_2,k_3,k_4) &=& N_1(k_1) N_2(k_2) \left[1+N_3 (k_3)\right]  \left[1+N_4 (k_4)\right] - (k_1 \leftrightarrow k_3 , k_2 \leftrightarrow k_4)  \, ,
\end{eqnarray}
where $N_i$ will depend on the particle in the process. We also assumed CP-invariance (which is true, since we work at tree level).

From the above phase space factors it is clear why collision rates are enhanced by the particle numbers in the initial and final states. For this reason the dominant process should be the one that involves only external $\gamma_+$.

\subsubsection*{Decays } \label{decays estimate}

In flat space the decay rate of a massive particle of mass $m$ with an axial coupling is simply given by $\Gamma_d \propto m_\phi^3/f^2$, so in the massless limit there should be no decay. Here however things are more complicated: we expect a nonzero decay (and inverse decay) rate due to the tachyonic instability of the $\gamma_+$ or, in other words, due to the energy transfer from the background to the fields.

We estimate such a decay rate by looking at the 1-loop correction to the 2-point function $\langle u_k u_k \rangle$, with gauge fields $\gamma_+$ running in the loop. Using the fact that the two point function is related to the mode functions via $\langle u_k u_{k'} \rangle = |u_k|^2 (2\pi)^3 \delta^3(\vec{k}-\vec{k'})$, we can relate the loop correction to an increase in the particle number and thus to an inverse decay rate\footnote{Note that using the Bogolyubov coefficients to define the particle number, as explained in sec.~\ref{bog}, one also finds that the two point function is related to the particle number as $\langle u_k u_{k} \rangle  \propto 1+ 2N_k + 2 \text{Re}\left[\beta_k \alpha_k^*\right]$ where $\alpha_k$ is the other Bogolyubov coefficient}.

To compute the loop correction we use the in-in formalism by considering the interaction Hamiltonian $H_\text{int} = \int d^3 x \sqrt{-g} \, \delta \phi \, F \tilde{F} /(4f) $. After Fourier transforming and integrating over the delta functions according to appendix \ref{Fourier transform} we get
\begin{eqnarray} \label{start}
\left< u_k u_k \right>_\text{loop} (\tau)&=& a^2 \left< \delta \phi_k \delta \phi_k \right>_\text{loop}  (\tau) \\
&=&a^2 \int_{-\infty}^{\tau} \frac{d \tau'}{(2\pi)^{12} (2f)^2} \int_{-\infty}^{\tau'} d \tau'' \int d^3 q \left| \vec{e}_+ (\vec{q}) \cdot \vec{e}_-(\vec{k}-\vec{q}) \right|^2\times  \nn \\
&& \times  \left< \left[A'_q (\tau') A_{|\vec{k}-\vec{q}|}(\tau') |\vec{k}-\vec{q}| \delta \phi_k(\tau') ,\left[A'_{|\vec{k}-\vec{q}|}(\tau'') A_q(\tau'')  q\, \delta \phi_k(\tau'') , \delta \phi_k (\tau)\delta \phi_k(\tau) \right]\right] \right>  + \nn \\ && + \, \text{perm.} \, , \nn
\end{eqnarray} 
where we omitted a factor of $\delta^3(0)$, $\vec{e}_{+,-}$ are the polarization vectors defined in \ref{pol ident} and we used $q\equiv|\vec{q}|$. Here [...] stands for a commutator and $\left< ...\right>$ is an expectation value. Then, by taking the time derivative of this expression and by making several approximations, described in  appendix~\ref{decays}, we arrive at
\begin{eqnarray} \label{e2} 
 &&\frac{dN_u(k)}{d\tau} \simeq - \frac{8 \xi b}{ (2\pi)^3 (2f)^2 a^2 k}  \times \\ 
 &&  \times \int \frac{dq \, q^3 |\vec{k}-\vec{q}|}{\text{min}(q,|\vec{k}-\vec{q}|)} N_u(k) (1+N_{\g_+}(q)) (1+N_{\g_+}(|\vec{k}-\vec{q}|))- N_{\g_+}(q) N_{\g_+}(|\vec{k}-\vec{q}|) (1+N_u(k)) \, ,  \nn
\end{eqnarray}
where $b$ is a number of order ${\cal O}(10^{-3})$ related to the angular integrals. Written in this form we can now identify the right hand side with the decay term $D^{+\phi}$ of eq.~(\ref{Boltzu}).

\subsection{Analytical estimates \label{estimates}}

Before proceeding to the numerical evaluation of the Boltzmann-like system of equations it is helpful to get some analytical estimates for the results. When solving numerically the system further assumptions will need to be made and so it is important to have some estimations to compare with. Of course we should keep in mind that the analytical estimations should be seen as order of magnitude estimates for $f$, and so $\xi$ could have $O(1)$ corrections.	
For the analytical approximation we will use the following assumptions:
\begin{itemize}
	\item 
	only subhorizon modes participate in the collisions, which is possible if $\xi \gtrsim 1$;
	\item 
	the collision integrals peak where the $N_{\g_+}$ is maximal, which means at energies $H\lesssim \omega \lesssim 2\xi H$;
	\item 
	thermalization happens when the collisions are faster than the Hubble expansion;
	\end{itemize}

For the gauge fields the dominant scatterings are $\g_+ \g_+ \leftrightarrow \g_+ \g_+$ because they are enhanced by the most powers of $N_{\g_+}$. After integrating the delta functions over the angles the scattering term has the form (we omit here factors of $a$): \small
\begin{eqnarray}
S^{++} = \frac{1}{\omega(k_1)}\int dk_2 dk_3 \frac{c_S}{ f^4} \left[  N_{\g_{+,3}} N_{\g_{+,4}} (1+N_{\g_{+,1}})(1+N_{\g_{+,2}}) - N_{\g_{+,1}} N_{\g_{+,2}} (1+N_{\g_{+,3}})(1+N_{\g_{+,4}})\right] \, ,
\label{s1}
\end{eqnarray} \normalsize
where $N_{\g_{+,i}}\equiv N_{\g_+}(k_i)$ and the coefficient $c_S$ is a dimension 4 combination of the energies $\omega(k_i)$ involved in the process. The scattering of $\g_+$ is only non-zero in the s-channel, whose matrix element is easy to compute and is given in eq.~(\ref{++++}).
In order to estimate when scatterings become relevant note that the Bose-Einstein factors in eq.~(\ref{s1}) can be also rewritten as:
\begin{eqnarray} 
N_{\g_{+,1}} N_{\g_{+,2}} +N_{\g_{+,1}}  N_{\g_{+,2}}  N_{\g_{+,3}} +N_{\g_{+,1}}  N_{\g_{+,4}}  N_{\g_{+,2}} - \nn \\ - N_{\g_{+,3}} \, ,  N_{\g_{+,4}}  N_{\g_{+,2}} -N_{\g_{+,1}}  N_{\g_{+,3}}  N_{\g_{+,4}} -N_{\g_{+,3}}  N_{\g_{+,4}} \, .
\end{eqnarray}
We assume the integrals to peak at energies $\omega_{i} \simeq \omega$, somewhere between $H$ and $2\xi H$, where the modes are more densely populated, with $N_{\g_{+,i}}\equiv N_{\g_+} \gg 1$. In this limit the above expression goes as $N_{\g_{+}}^3$, hence, we estimate 
\begin{eqnarray}S^{++}\approx  \frac{\omega^5}{\beta_S f^4} N_{\g_+}^3 \, .\label{scattest}
\end{eqnarray}
Here $\beta_S = {\cal O}(10^4)$ is a numerical factor that comes from the collision integrals and the cross sections. The scatterings become relevant when this term is comparable with the left-hand side of eq.~(\ref{BoltzA}) which is of order $N_{\g_+} H$ (or than the source term, which is of the same order). Therefore, the condition in the particle number to have thermalization is
\begin{eqnarray} N_{\g_+} \gg  \sqrt{  \beta_S \frac{H f^4}{\omega^5}} \, . \label{C1}
\end{eqnarray}
In the next section we compare this estimate with the numerical results. This can also be translated into a bound on $\xi$, for a given $f/H$, by using the fact that, in absence of thermalization, the particle number depends exponentially on $\xi$. We can estimate the total particle number in the band $1<-k\tau<2 \xi$ by numerically integrating $N_{\gamma_+}\approx 2 k |A_+|^2$, using the exact mode functions  and fitting with an exponential. This yields $N_{\g_+} \approx 10^{-4}e^{4.5 \xi}$. Using this expression in the above condition for thermalization, and setting $\omega\approx H$, then gives
\begin{eqnarray}
 \xi \gtrsim  0.45 \ln \left(\frac{f}{H} \right) +2.7  \, ,
 \label{thermalcondition}
\end{eqnarray}
which we will compare with numerical results. From this estimate we see that, even if the scattering efficiency is suppressed when $f\gg H$, thermalization can still happen for sufficiently large $\xi$.

Now we apply the same type of analysis for the decays, which have the form
\begin{eqnarray}
D^{+\phi} &=& \frac{1}{\omega_1}\int dk_2  \frac{c_D}{ f^2} \left[   N_{u_3} (1+N_{\g_{+,1}})(1+N_{\g_{+,2}}) - N_{\g_{+,1}} N_{\g_{+,2}} (1+N_{u_3}) ]  \right) \, ,
\end{eqnarray}
where $c_D$ is a dimension 3 coefficient. Similarly to the scatterings, the decay term can be approximated by
\begin{eqnarray}
D \approx  \frac{\omega^3}{\beta_D f^2} N^2_{\gamma_+} \, ,
\end{eqnarray}
where we assumed $N_{\g_+}\gg N_\phi$. Here $\beta_D \approx 10^5/\xi$ is the inverse of the pre-factor of eq.~\ref{e2}.
Comparing $D$ with $N_{\g_+} H$ gives
\begin{eqnarray} \label{C2}
N_{\g_+}(\omega) \gg \beta_D \frac{H f^2}{\omega^3} \qquad \implies \qquad \xi\gtrsim 0.45 \ln \left(\frac{f}{H}\right)+4.3 \, ,
\label{CD}
\end{eqnarray}
which shows that they are subdominant with respect to the scatterings, eq.~(\ref{thermalcondition}). So there may be a regime in which scatterings are in equilibrium but decays are not, which will lead to the presence of a chemical potential, as we will discuss below.

Finally we can estimate the threshold of thermalization for $\phi$ and $\g_-$. In the case of $\g_-$ the collision terms can be estimated as $S^{+-}\approx  \omega^5 N_{\g_+}^2 N_{\g_-}/ (\beta_{-} f^4)$, with $\beta_{-}={\cal O}(10^2)$, which should be compared to the left-hand side of the  Boltzmann equation, of order $N_{\g_-} H$.
This leads to the condition
\begin{eqnarray} N_{\g_+} \gg  \sqrt{  \beta_- \frac{H f^4}{\omega^5}} \qquad \implies \qquad \xi\gtrsim 0.45 \ln \left(\frac{f}{H}\right)+2.2 \, , \label{Cminus}
\end{eqnarray}
which is actually realized even more easily than  eq.~(\ref{C1}), since $\beta_-\ll \beta_S$.
The case of $\phi$ is also analogous and leads to
\begin{eqnarray} N_{\g_+} \gg  \sqrt{  \beta_\phi \frac{H f^4}{\omega^5}} \qquad \implies \qquad  \xi\gtrsim  0.45 \ln \left(\frac{f}{H}\right)+2.4 \, , \label{Cphi}
\end{eqnarray}
where $\beta_{\phi}\approx 5 \cdot 10^2$ and so it is also realized almost at the same time as the above eq.~(\ref{Cminus}).

Among the above reactions all the scatterings conserve particle number and so they can lead to kinetic equilibrium, but possibly with a nonzero chemical potential $\mu$; only the decays can change the particle number and lead to chemical equilibrium, driving $\mu$ to 0 (we are not considering here other processes such as $2\leftrightarrow 4$ scatterings, which could also do the same). Therefore, we expect the following behavior if, for a fixed value of $\xi$, we decrease $f/H$: (1) first eqs.(\ref{Cminus}) and (\ref{Cphi}) would be satisfied, which means that $N_{\g_-}$ and $N_{\phi}$ should start tracking $N_{\g_+}$; (2) then $\g_{+}$ should reach kinetic equilibrium when condition~(\ref{C1}) is fulfilled, so that all species should reach a Bose-Einstein distribution, possibly with a nonzero $\mu$; (3) finally also the decays go in equilibrium driving $\mu$ to zero, reaching blackbody distributions.

Such a picture would be modified in presence of other interactions (as in the case of SM interactions): in this case gauge fields can reach both kinetic and chemical equilibrium more easily, because of fast $2\leftrightarrow 2$ processes, as well as number changing processes. However, this might not happen to $\phi$, if its only interaction is the axial coupling, and we may have the situation in which gauge fields are fully in equilibrium, but not $\phi$.

\subsection{Standard Model couplings}

So far we have only considered  processes involving the axial coupling between $\phi$ and the gauge fields, which are the only relevant ones if the gauge field is a ``dark photon", not belonging to the SM.
In this section we consider the case in which the gauge field in the axial coupling belongs instead to the Standard Model (SM), either the U(1) hypercharge or a non-abelian SU(2) or SU(3) gauge boson~\footnote{Depending on the couplings of the Higgs to gravity and in a very high energy regime the Higgs vev could be zero and so all gauge bosons could be massless during inflation, otherwise one can adapt these estimates to other cases, where for instance only photons and gluons are massless.}. In fact this is the most interesting situation for two reasons: first it gives a natural way to reheat the universe into SM particles and second such a coupling of the inflaton to the SM may make the model directly testable. 

The purpose of this section is to study whether the SM self-interactions and scatterings involving gauge bosons and SM charged fields could help in thermalizing the system. 
The interesting feature of such interactions is that they are not suppressed by powers of $1/f$ like the previous diagrams. Instead, they are proportional to gauge coupling constants and so, if $f\gg H$, they are the relevant interactions for thermalization. Moreover, this makes the scenario more predictive since there is only one parameter ($\xi$) that sets the thermalization condition for the gauge bosons. 
However, as we mentioned before, in this case the $\phi$ perturbations might not be thermalized, depending on the value of $f/H$.

We will not include the SM interactions in the numerical evaluation of the Boltzmann-like equations, in the next section, since the purpose of this work is just a proof of principle for thermalization. We leave a deeper and more precise study with all such interactions for future work. Nevertheless, we can still estimate the onset of thermalization due to scatterings with SM particles and compare it with the interactions generated from the axial coupling.

\subsubsection*{Photon-Fermion scatterings}

Let us first consider gauge boson-fermion scatterings, through pair production. We consider the case of electron-positron production by photon scattering where the differential cross section is given, in the center-of-mass frame and in the high energy limit, by
\begin{eqnarray}
\frac{d \sigma}{d \Omega}_{\gamma \gamma \rightarrow e^- e^+} = \frac{2 \pi \alpha^2}{(4\omega)^2} \left(\frac{1+\cos^2\theta}{\sin^2 \theta}\right) \, ,
\end{eqnarray}
where $\alpha= e^2/ (4\pi)$, $e$ is the electric charge and  $2\omega$ is the center-of-mass energy. The total cross section has a $\log $ divergence for small angles which can be regulated by imposing an IR cutoff, in our case given by the Hubble scale $H$, or, in the case of thermalization, by a thermal mass of order $e T$. 
Now we can compare this expression with photon-photon scattering mediated by $\phi$, which in the center-of-mass frame is 
\begin{eqnarray}
\sigma_{\gamma \gamma \rightarrow \gamma \gamma}^\phi= \frac{|M|^2}{64\pi^2 (2\omega)^2} \simeq \frac{1}{4} \left(\frac{\omega}{f}\right)^4  \frac{1}{64\pi^2 (2\omega)^2} \, .
\end{eqnarray}
Comparing the two cross sections we have 
\begin{eqnarray}
  \frac{\sigma_{\gamma \gamma \rightarrow e^- e^+}}{\sigma^\phi_{\gamma \gamma \rightarrow \gamma \gamma}} \simeq \frac{ \frac{2 \pi \alpha^2}{(4 \omega)^2} } { \frac{1}{4} \left(\frac{\omega}{f}\right)^4  \frac{1}{64\pi^2 (2\omega)^2}} = 128 \pi ^3 \alpha ^2 \frac{f^4}{\omega^4} \, ,
\end{eqnarray}
where we assumed the logarithmic term to give a coefficient of order one and the typical energy of the process to be $\omega$. It is clear that, if $f \gg H$, the cross section for pair production is much larger than the scatterings mediated by $\phi$. In particular if we extrapolate the Standard Model to high energies the relevant coupling is the U(1) hypercharge, whose $\alpha$ changes from $1/40$ at $10^{14}$ GeV to $1/60$ at $10^2$ GeV. Therefore, the ratio of cross-sections is always above one for any $f>\omega$, which is anyway required for the validity of effective field theory. 

On the other hand, pair production is less Bose-enhanced than photon-photon scattering by one power of $N_{\g_+}$. To estimate the threshold for thermalization we compare $H$ with $n_{\g_+} \sigma_{\gamma \gamma \rightarrow e^- e^+} $ where $n_{\g_+} \approx  N_{\g_+} H^3$ is roughly the particle number density. Therefore, the thermalization condition is
\begin{eqnarray}
N_{\g_+} \gg  \frac{8}{\pi \alpha^2}  \,,
\end{eqnarray}
where we again assumed $\omega\simeq H$. Note that this is quite rough, since we have neglected powers of $\xi$ in $\omega$ and $n_{\g_+}$.
Moreover, we have just considered one diagram, while in reality we have to also sum over all particle-antiparticle pairs of charged fermions in the standard model. Having this is mind and taking into account fractional charges we get an extra factor of approximately\footnote{Here we only consider pair production terms, while adding also Compton scatterings probably increases the result by roughly another factor of 2, at large occupation numbers.} 85. Compared to eq.~(\ref{C1}), this means that thermalization will happen faster through SM interactions than through the axial coupling for any value of $f>H$. In particular, using the value of $N_{\g_+}$ derived in the previous subsection, $N_{\g_+} = 10^{-4} e^{4.5 \xi}$, and the value of $\alpha$ at $10^{14}$ GeV the system would thermalize when $\xi \gtrsim 2.9$.

We can also compare these scattering rates with the rate of particle production by the Schwinger effect given by $\Gamma_\text{S} \simeq \frac{e^2 E^2}{6 \pi^3 H^3} \exp\left( {\frac{-\pi m_e^2}{e E}} \right)$~\cite{Schwinger:1951nm, Hayashinaka:2016qqn, Tangarife:2017rgl}, where $E$ is the coherent electric field and $m_e$ is the electron (or any charged fermion) mass. If we estimate the electric field as $E^2 \simeq N_{\g_{+}} H^4$ and consider the regime $E \gg m^2_e$, we find a similar threshold for thermalization in terms of $\xi$.

\subsubsection*{Gauge-field self interactions}

Another case in which thermalization of gauge fields can happen quite easily is if the gauge bosons belong to $SU(2)$ or $SU(3)$.
In this case, there will be cubic and/or quartic self-interactions already at tree-level, which are boosted in the Boltzmann equation by the number density of positive helicity gauge bosons. Consider for example a quartic tree level interaction, with  cross section $\propto g_s^2$, with $g_s$ the gauge group coupling constant. The cross-section of gluon-gluon scattering is\footnote{We neglect factors of order 1 in the total cross-section.} 
\begin{eqnarray}
\sigma_{gg\rightarrow gg } \simeq \frac{9 \pi \alpha_s}{2 (4 \omega)^2} \, ,
\end{eqnarray}
where $\alpha_s= g_s^2/ (4\pi)$. Thus, the condition for thermalization is
\begin{eqnarray}
N_{\g_+}  \gg  \frac{1 }{\alpha_s} \, .
\end{eqnarray}
For $SU(3)$, whose coupling constant $\alpha_s=g_s^2/(4\pi)$ runs from $1/40$ at $10^{14}$ GeV to $\simeq 1$ at GeV scale \footnote{Note that this analysis is only valid for gauge-fields in the perturbative regime where the propagator and the equation of motion remains unmodified by self-interactions.}, this means that gluon scatterings are the most efficient process for thermalization and the system would thermalize also around $\xi \simeq 2.9$.

\section{Numerics}  \label{numerics}

This section is devoted to the numerical evaluation of the Boltzmann-like system of equations described in the previous section. Due to the several approximations we have used, the numerical results do not aim at being precise but instead at giving a rough description of the phenomena. Therefore, the results should be seen as an order of magnitude estimate. We start by listing the main approximations and simplifications used:
\begin{itemize}
	\item
	We work with a finite set of momenta. For this reason we need to discretize the $k$-integrals appearing in both the scatterings and decays terms in eqs.~(\ref{BoltzA}), (\ref{BoltzA2}) and (\ref{Boltzu}). Therefore, we substitute
	\begin{eqnarray}
	\int dk \rightarrow \Delta k \sum_k  \nn \, ,
	\end{eqnarray}
	where $\Delta k$ is the mode spacing. We typically use 10 discrete modes, multiples of a given $k_\text{min}$.
	
	\item 
	We only consider subhorizon modes, where the flat space result and the particle interpretation are a good approximation. We start our simulation when the largest mode $k_\text{min}$ is subhorizon, using $|k_\text{min} \tau|=2$,  and stop it when such mode goes superhorizon ($|k_\text{min} \tau| < 1$). 	
	
	\item
	We give as initial condition for $\g_+$ the solution of the equation of motion eq.~(\ref{EOM for A}). For $\g_-$ and $\phi$ one can use the vacuum solution and eq~(\ref{solu}), respectively, although the final results will basically be insensitive to such choice. 
	
	\item
	To estimate the  temperature of the system at thermalization, $\bar{T}$, one would need to ensure that the chosen window of modes covers the bulk of the energy, which consists of modes of momenta $k/a \lesssim \bar{T}$. However, $\bar{T}$ grows exponentially with $\xi$ and the instability band consists of modes of $k/a$ between $H$ and $\xi H$. Thus, for large $\xi$ we would need a very large window of momenta to extract correctly the temperature. For numerical reasons we do not consider such a large box but only a band of modes between $H$ and ${\cal O} (10)  \,H$ approximately. For this reason we cannot  evaluate correctly $\bar{T}$ when $\xi$ is large, although the system still approaches a Bose-Einstein distribution in the chosen window.
	
	\item
	We neglect any backreaction on the scalar field equation of motion. This would be relevant when $\langle F\tilde{F}\rangle/f$ becomes of order of the derivative of the potential $dV/d\phi$ and would require to follow the treatment of~\cite{Anber2009,Notari2016} and to include in the system the background equation for $\phi$. We postpone this richer situation to future work.
	
	\item
	As already stressed in section \ref{sect:Boltzmann}, we assume the functions $g_{u,A}$, defined in eqs.~(\ref{eqsnumbers}) and (\ref{eqsnumbers2}), to be always given by the solution of the equations of motion in absence of collisions. This assumption might fail after thermalization, but should be valid until there.
	
	\item The expressions for the scatterings are based on flat space cross sections, which is justified if thermalization is faster than the Hubble expansion. The decays are instead based on translating results from the two-point function of $\delta\phi$ in the in-in formalism. A more rigorous treatment that combines non-equilibrium field theory with Boltzmann equations is of course desirable, but we think that our treatment should give a correct picture and a reliable order of magnitude estimate. 
	
	Note that we use a massless dispersion relation for all the species we consider. This implies that we will get massless BE distributions, while in reality the gauge fields have nontrivial dispersion relations due to the axial coupling. In fact, as we already stressed in sec.~\ref{thermalization}, we expect deviations from this limit for modes such that $k/a\lesssim 2 \xi H$, but we postpone the analysis of such deviations to a forthcoming publication.
	
	\item Our criterion for thermalization is to verify if the average difference to a BE distribution is smaller than $1\%$. The average difference is defined as
	\begin{eqnarray} \label{avdiff}
	\frac{\Delta N}{N} \equiv \frac{1}{N_\text{tot}}  \sum_k \frac{N^\text{norm} (k)-N^\text{eq}(k,T)}{N^\text{eq}(k,T)} \, ,
	\end{eqnarray}
	where $N_\text{tot}$ is the total number of modes, $N^\text{eq}(k,T)$ is the Bose-Einstein distribution computed at a temperature $T$ extracted from the energy density of the system and $N^\text{norm} (k)$ is the particle number normalized by the ratio $N^\text{eq}(k_*,T)/N(k_*,T)$ for one given value of $k_*$. The reason to use $N^\text{norm} (k)$ is that, for large $\xi$, the mode $k \approx T$ is outside the box which means that the estimated temperature is not accurate and so there would be an offset between the equilibrium distribution and the numerical result. Note that if the system thermalizes but decays are not efficient the distribution would have a chemical potential ($\mu$), since particle number is conserved at the time of thermalization. We find that the chemical potential is small compared to the temperature and it is only visible in the smallest $k$ mode for such cases (see appendix \ref{numbdist}). For this reason we discard the first mode in the average difference.

	\item Some of the plots include regions with $f \simeq H$, which are close to the UV cutoff. However, here we only want to show that our estimates agree well with the numerical results in a region of parameter space where the numerics works well, so that we can then extrapolate for larger values of $f$ and $\xi$, where the numerical treatment becomes more difficult to perform.
	
\end{itemize}

\begin{figure} 
	\centering
	\includegraphics[scale=0.285]{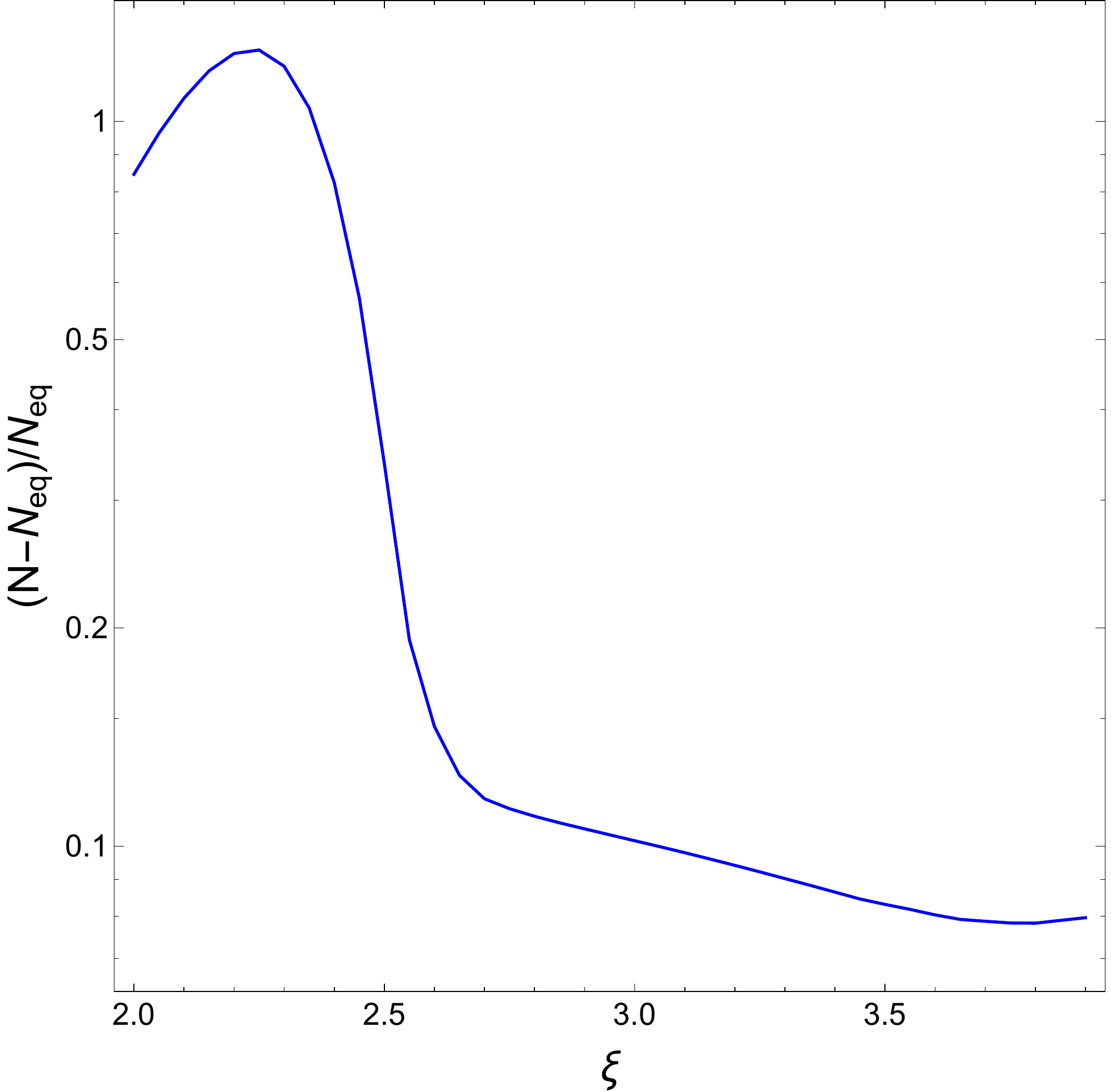}
	\includegraphics[scale=0.4]{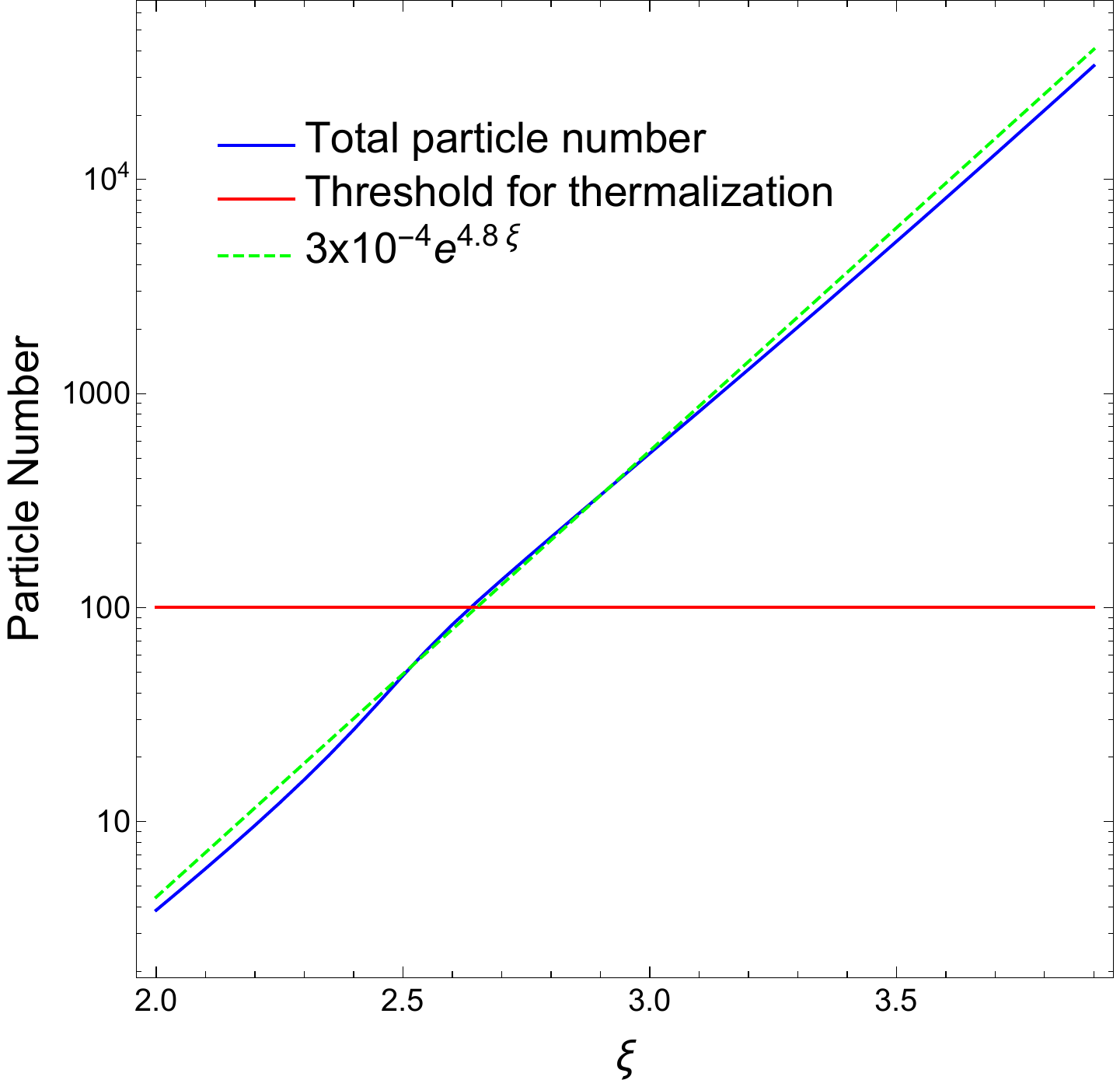}
	\includegraphics[scale=0.285]{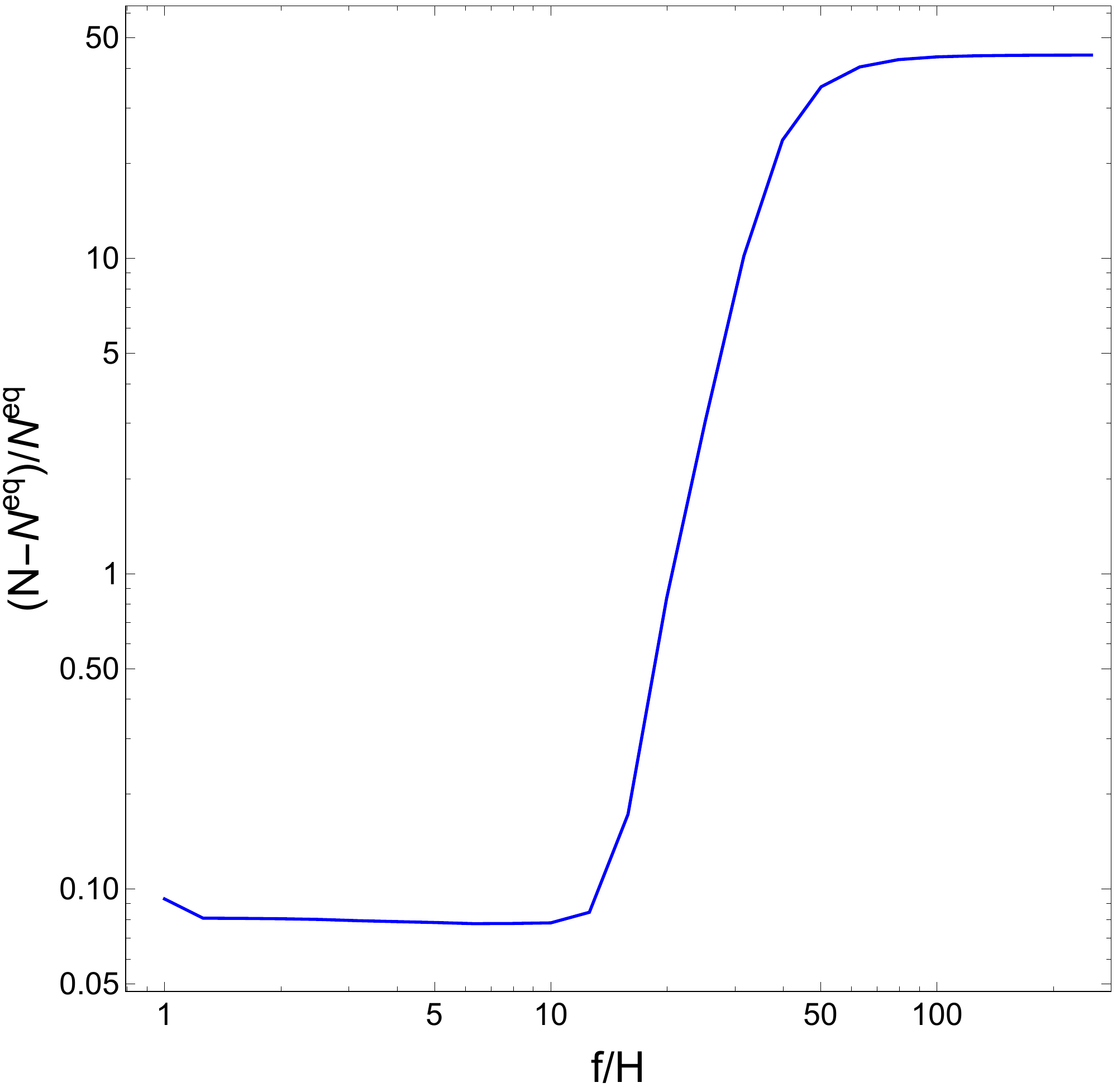}
	\includegraphics[scale=0.4]{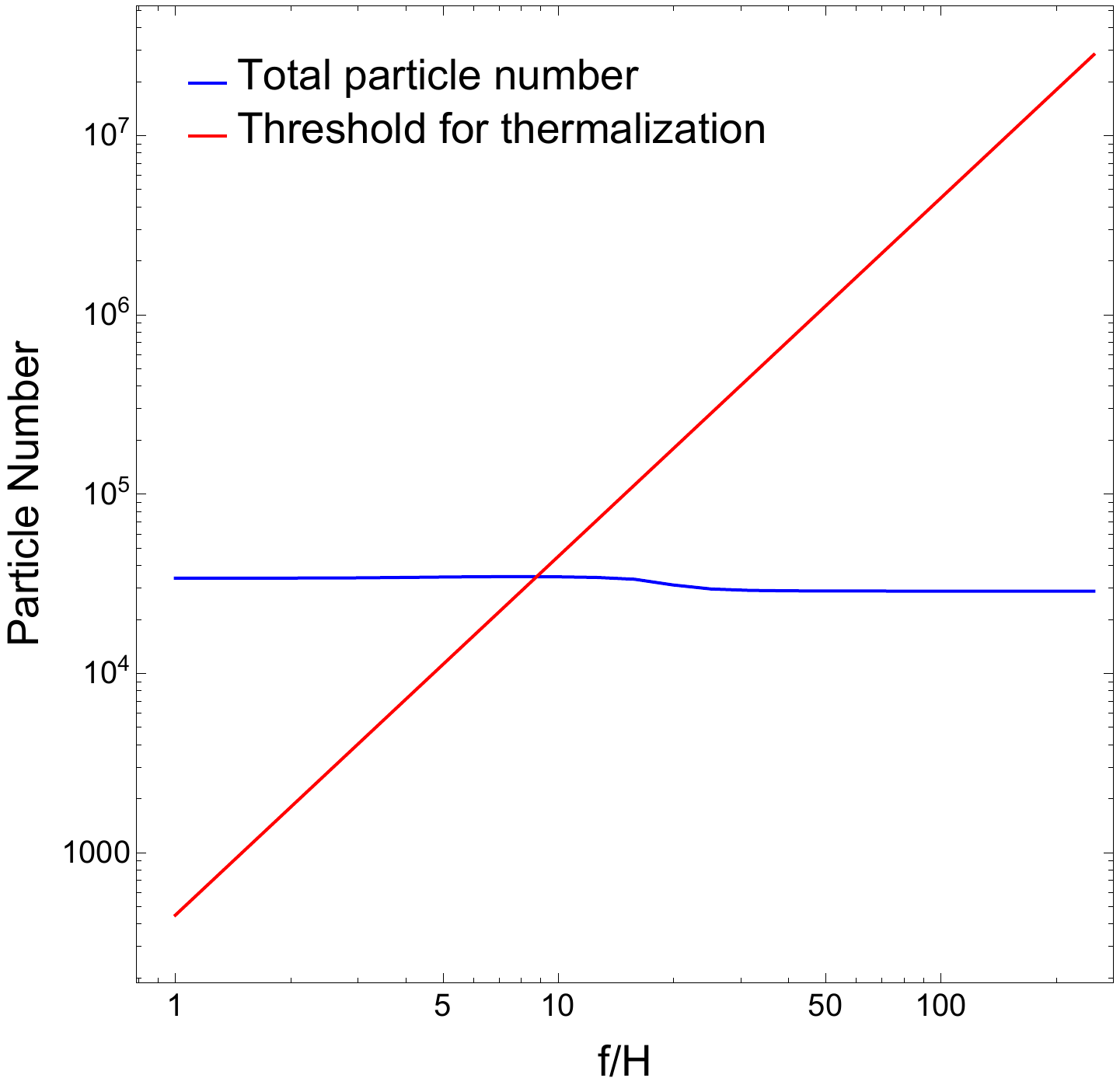}
	\caption{ \label{A+fig}Left plots: average normalized difference to a Bose-Einstein distribution (defined in eq.~(\ref{avdiff})). Right plots: Total particle numbers and threshold for thermalization. Upper plots: $f=H$ and varying $\xi$. Bottom plots: $\xi=3.9$ and varying $f/H$.}
\end{figure}

On generic grounds, the numerical results confirm our expectations, {\it i.e.}, that thermalization occurs when the particle number, controlled by $\xi$, is larger than a given threshold. If $f/H$ decreases, the threshold becomes lower and the system thermalizes more easily. The thresholds for thermalization are in agreement with eq.~(\ref{thermalcondition}). 
We separate the results in 2 different cases of interest. First we consider a system with $\gamma_+$ modes only. This is the case where thermalization is the most efficient because the gauge field number is larger. Then, we include scatterings with $\g_-$ and $\phi$ and numerically evaluate the threshold for thermalization. In appendix \ref{numbdist} we also add some plots which show the complete distributions as a function of momenta, at a given time.

Let us start by considering only $\gamma_+$ gauge fields and their scatterings, with matrix elements given in eq.~(\ref{++++}). In fig.~\ref{A+fig} we show the total particle number and the average difference to a thermal distribution $\Delta N/N$, defined in eq.~(\ref{avdiff}), for fixed $f$ and varying $\xi$ and vice-versa, evaluated at the end of the simulation (when the longest mode is horizon size). 
We assume the system has thermalized when $\Delta N/N \lesssim 0.1$. When $\xi=3.9$, $\Delta N/N$ decreases sharply to roughly $0.1$ for $f\lesssim20H$, while for $f=H$ that happens for $\xi\gtrsim 2.7$. 
This is in agreement with the plots on the right which show that the analytical estimate for the threshold of thermalization, using eq.~(\ref{C1}), intersects the total particle number $N^\text{total}_{\gamma_+}=\sum_k N_{\gamma_+}(k)$ at roughly the same values of $f$ and $\xi$.
In the upper right plot we also see that the particle number depends exponentially on $\xi$ and is well approximated by $N_{\g_+}\simeq 3  \times 10^{-4} e^{4.8\xi}$, as expected from the analytical estimates in sec.~\ref{sect:Boltzmann}. 

\begin{figure} 
	\centering
	\includegraphics[scale=0.4]{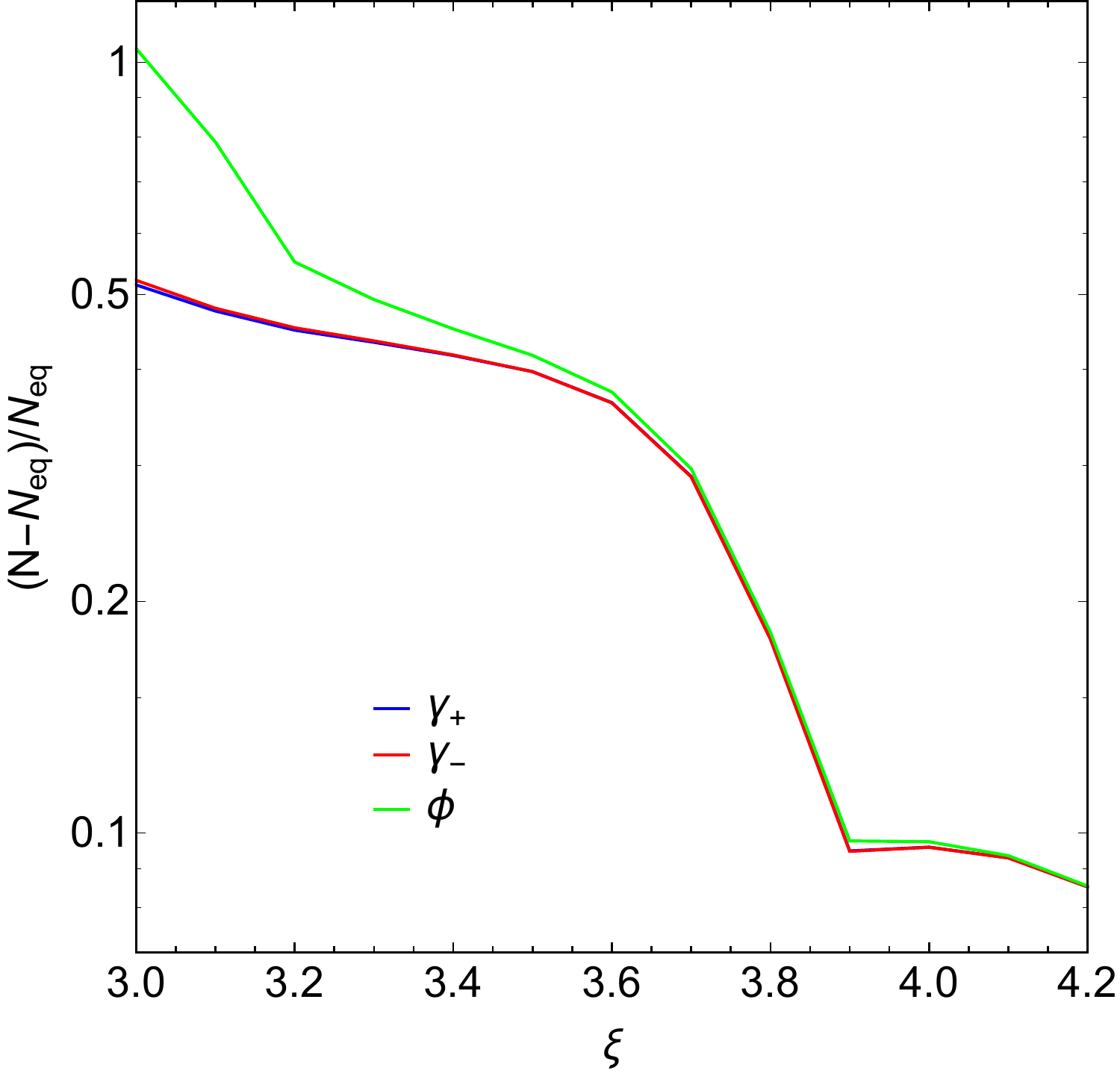}
	\includegraphics[scale=0.425]{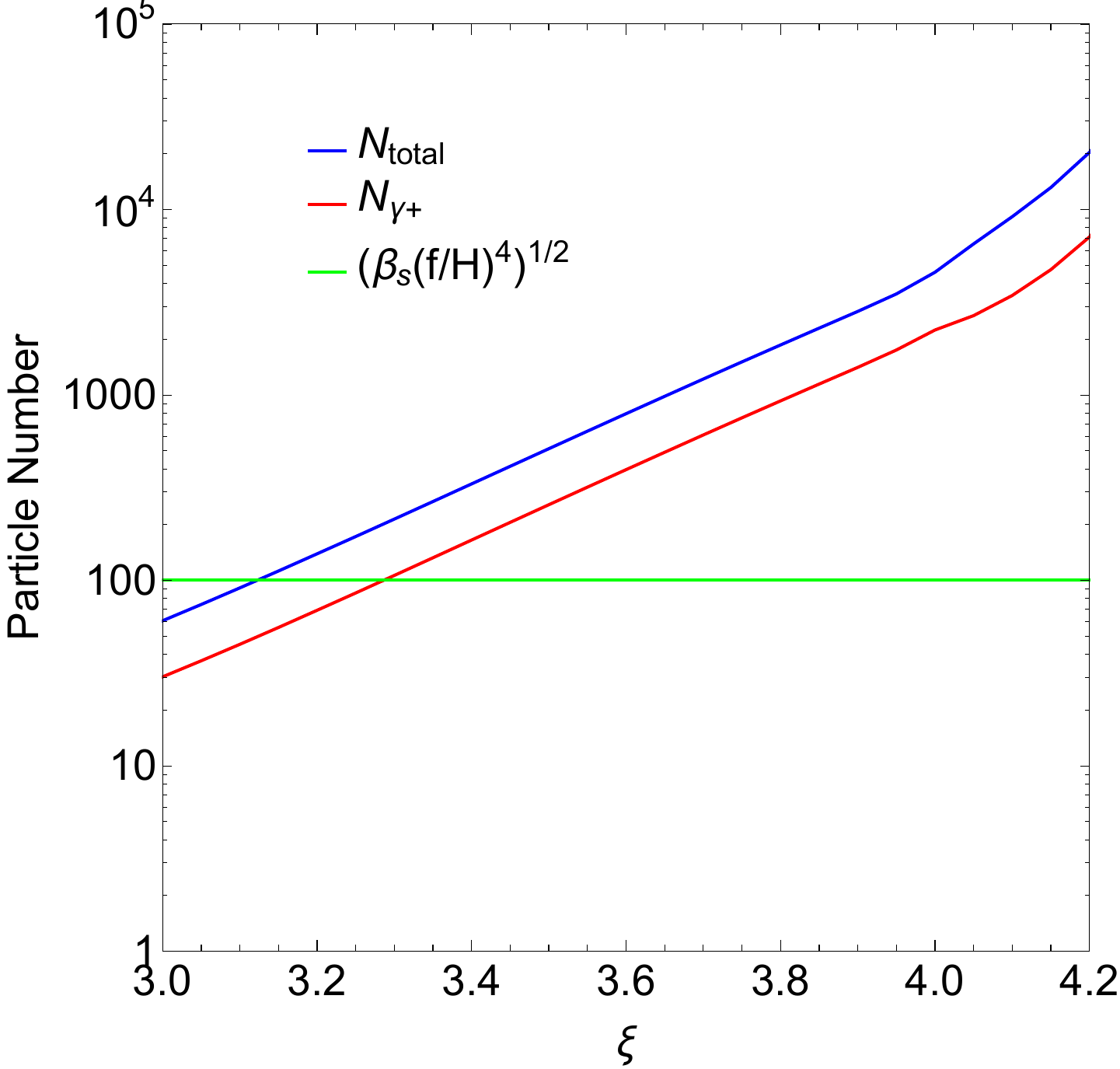}
	\includegraphics[scale=0.3]{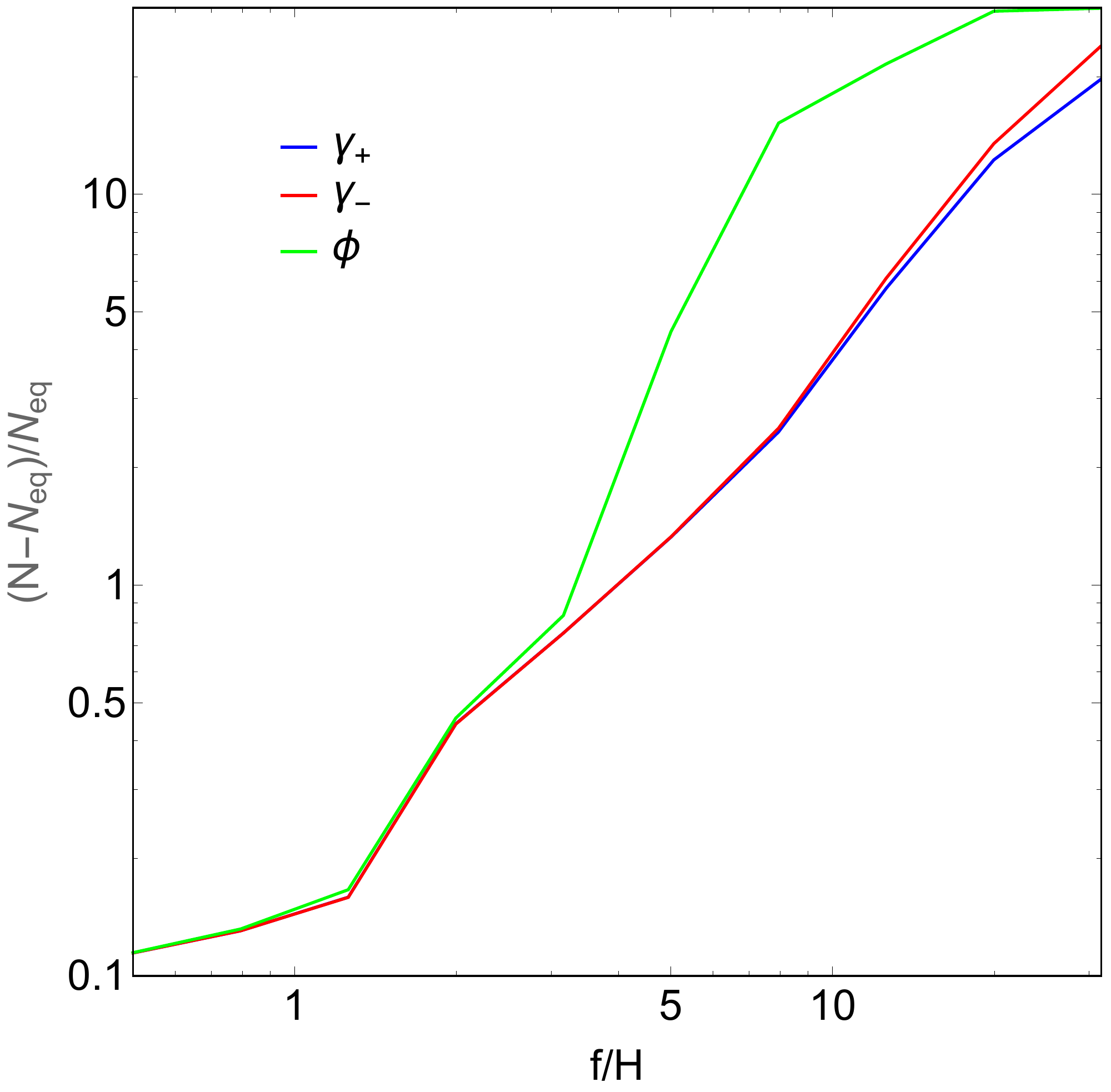}
	\includegraphics[scale=0.31]{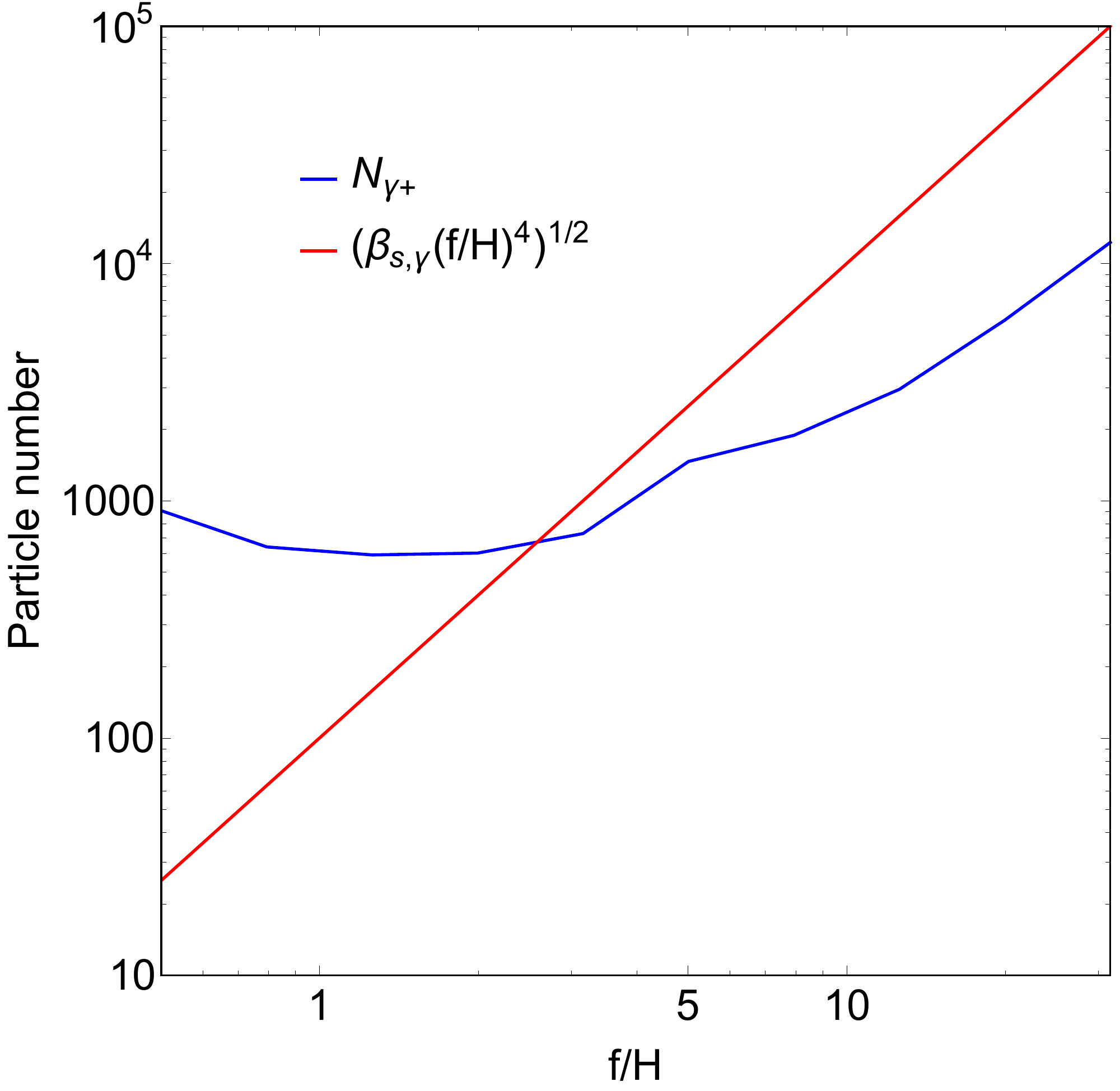}
	\caption{\label{A+-phi} Left plots: average normalized difference to a Bose-Einstein distribution (defined in eq.~(\ref{avdiff})) for each of the degrees of freedom ($\phi, \gamma_+, \gamma_-$). Right plots: Total particle numbers and threshold for thermalization. Upper plots: $f=H$ and varying $\xi$. Bottom plots: $\xi=3.9$ and varying $f$.}
\end{figure}

In the second case we consider all scatterings and decays involving $\phi, \gamma_+$ and $\g_-$. The presence of decays and of more degrees of freedom leads to a decrease in the $\gamma_+$ particle number which implies less efficient thermalization, {\it i.e}, it happens for smaller (larger) value of $f$ ($\xi$).  In fig.~\ref{A+-phi} we show the same plots as in fig.~\ref{A+fig}. By looking at the average differences to a Bose-Einstein distribution (left plots) one can see that for $\xi=3.9$, lower plots, the two gauge field polarizations thermalize at $f\lesssim 4 H$ while for $f=H$ all species thermalize when $\xi\gtrsim 3.9$. Comparing the average differences, on the left, with the particle numbers and estimates, on the right, we see that thermalization requires a value of $\xi$ about 0.6 larger and a value of $f/H$ about two times lower than the estimate.

Such results seems to have parallel in the results obtained in \cite{Adshead:2016iae} where the authors also found that the power spectrum of both polarizations of the gauge field and of the inflaton perturbations follow each other while, at the same time, there is a transfer of energy to the UV (fig. 7 of  \cite{Adshead:2016iae}).

Finally, we also evaluate numerically the threshold for thermalization by collecting the values of $\xi$ and $f$ such that $\Delta N/N<0.1$. The threshold is well fitted by
\begin{eqnarray}
\xi = 0.44 \log \left(\frac{f}{H}\right) + 3.4 \, ,
\end{eqnarray}
which is close to the estimated threshold in eq.~(\ref{thermalcondition}), up to an offset of 0.7 in $\xi$. We will use this threshold equation in the next section when discussing the phenomenology.

\section{Phenomenology}  \label{phenomenology}

After thermalization is reached, however, the evolution is expected to be non-trivial. Indeed the temperature should decrease and the system could also depart from a thermal spectrum at low momenta. In fact, having a temperature normally generates thermal masses for the gauge fields, which tend to screen the instability and, as a result, the source will be less efficient and the temperature is expected to decrease. Moreover, the energy density $\rho_{\phi}$ extracted from the scalar field is given by $d\rho_\phi/dt=\dot{\phi}/(4f) \langle  F \tilde{F} \rangle=\dot{ \phi}/f \int (d^3k)/(2 \pi)^3 k d/d\tau(|A_+|^2-|A_-|^2) /a^4$, averaged over a thermal distribution; such a quantity should be suppressed if $\gamma_+$ and $\gamma_-$ are in equilibrium, since the scatterings tend to restore parity symmetry. We do not try to estimate such quantity exactly here because, while $N_{\g_-}$ is easy to compute in equilibrium, the same is not true for $N_{\g_+}$, due to its complex frequency $\omega_+$. We postpone to future work a refined numerical treatment, including such effects, but we anticipate in this section some of the expected features,  under the assumption that  thermalization is successful.  Generically we can anticipate how the scalar and tensor curvature perturbation, $\zeta$ and $h$ respectively, should be affected by thermalization:

\begin{itemize}
	
	\item {\it Superhorizon conservation:}

On superhorizon scales the gauge field mode function goes to a constant, so its energy density decreases as $a^{-4}$. Because the sourcing of adiabatic curvature perturbation is proportional to this quantity, we expect a negligible isocurvature sourcing of the scalar curvature perturbation in the uniform density gauge, $\zeta$. In appendix \ref{iso} we elaborate on this point.  Therefore, it should be a good estimate to evaluate correlators of $\zeta$ at horizon crossing.  We also  disregard here the possibility that a backreaction regime, which could change the predictions on perturbations, can be present at any time during inflation.
	
	\item  {\it Loop corrections:}
	
	In the absence of thermalization strong constraints on $\xi$ were derived from non-Gaussianities~\cite{Barnaby2011, Bartolo2014, Ferreira2014a} and the requirement of perturbativity \cite{Ferreira2015a, Peloso2016} due to loop corrections involving the gauge fields. We expect these corrections to become smaller in the thermal regime by the fact that energy moves from the horizon size to smaller scales, possibly $k/a \simeq T$, and so at horizon crossing the effect on correlators should be  smaller.
	
	\item  {\it Parity symmetry:}
	
	On top of the above suppression, $ \langle F \tilde{F} \rangle $ should also be suppressed in a thermal environment due the tendency of the scatterings to restore parity symmetry. Because each vertex introduces one power of $F \tilde{F}$, loop corrections from the axial coupling to odd $\zeta$ correlators are proportional to $ \langle F \tilde{F} \rangle $ and so they will also be suppressed.

\end{itemize}

\subsection{Power spectrum and tensor to scalar ratio}

In this subsection we estimate the power spectrum of curvature perturbation assuming that a thermal regime is reached and under the assumptions stated above. We will have different cases depending on whether perturbations are thermalized or not. 

Let us briefly review, first, the standard vacuum case. In this case the mode functions associated with the canonically normalized field are given, in the de-Sitter limit, by
\begin{eqnarray}
\left|u_k \right|^2 = \frac{1}{2 k} \left|1-\frac{i}{k\tau}\right|^2 \underset{-k\tau \rightarrow 0}{\simeq}\, \frac{1}{2k^3 \tau^2} \, ,
\end{eqnarray}
which means that the power spectrum of the scalar curvature perturbation in the uniform density gauge, $\zeta$, in the superhorizon limit, $-k\tau \rightarrow 0$, is~\cite{Brandenberger1993}
\begin{eqnarray} 
P^\text{vac}_\zeta \equiv \frac{|\zeta_k|^2 k^3}{2 \pi^2}= \left|\frac{H k^3}{2\pi^2 a \dot{\phi} } u_k\right|^2 =  \frac{H^4}{4\pi^2 \dot{\phi}^2 } \, .
\end{eqnarray}
Alternatively, one could have evaluated the subhorizon expression for $u$ at horizon crossing, $-k\tau \simeq1$, and plugged it into the definition of $\zeta$, which is constant on superhorizon:
\begin{eqnarray}
\left|u_k \right|^2 \underset{-k\tau \rightarrow 1}{\simeq}\, \frac{1}{2k} \qquad \Rightarrow \qquad  P^\text{vac}_\zeta= \left. \frac{H^2 k^3}{2 \pi^2 \dot{\phi}^2} \frac{1}{ 2 k a^2}\right\vert_{-k\tau \rightarrow 1}=\frac{H_*^4 }{4 \pi^2 \dot{\phi}_*^2} \, ,
\end{eqnarray}
where $*$ denotes quantities evaluated at horizon crossing. 
This last result is actually more general because $H_*$ and $\dot{\phi}_*$ are evaluated at horizon crossing and so it also holds in the quasi de Sitter case.

If the inflaton perturbations are not thermalized they will still be given at zero order by the above equations, but they will be sourced at one-loop by thermal gauge field fluctuations. This case is, however, more complicated and we also postpone it to a future analysis.

If the inflaton perturbations are instead thermal, the expected outcome is much simpler by following the same strategy of evaluating $u_k$ at horizon crossing. In this case $u_k$ is related to the particle number $N_k$ through eq.~(\ref{number}), which should be reliable slightly inside the horizon. Then, using $|u'_k|^2 \simeq k |u_k|^2$, we have
\begin{eqnarray}
\left|u^\text{therm}_k \right|^2 = \frac{1}{k}   \left. \left(\frac{1}{2}+N_k \right) \right\vert_{-k\tau  \rightarrow 1}{\simeq}  \,  \frac 1 k \frac{T_*}{H_*} \, ,
\end{eqnarray}
which then implies that the curvature perturbation is simply obtained by replacing the vacuum particle number (1/2) with the thermal value $T_*/H_*$: 
\begin{eqnarray}
P^\text{therm}_\zeta = \frac{T_*}{H_*}  \frac{H^4_*}{2 \pi^2 \dot{\phi}_*^2 } \, . \label{PzetaT}
\end{eqnarray}
Note that this prediction naturally has similarities with those of warm inflation \cite{Berera2008, Bartrum:2013fia} from the fact that both use the thermal particle number. However we do not consider any friction term, induced by the thermal bath, in the equation of motion for $\zeta$.
Regarding the tilt of the spectrum, although both cases give a nearly scale invariant power spectrum, in the thermal case the departure from scale invariance has a different functional dependence on the slow-roll parameters. In the vacuum case the spectral index is due to the time variation of $H$ and $\dot{\phi}$ at horizon crossing, while here the ratio $T/H$ at horizon crossing also matters.

Indeed the spectral index is given by:
\begin{eqnarray}
n_s-1\equiv \frac{d \ln P^\text{therm}_\zeta }{d \ln k} &=& -6 \epsilon_H+2\eta + \frac{d \ln(T_*/H_*) }{d \ln k}  \, ,
\end{eqnarray}
where we have used the following definitions for slow-roll parameters:
\begin{eqnarray}
\epsilon_H \equiv -\frac{\dot{H}}{H^2} \, , \qquad \, 
\eta\equiv \epsilon_H- \frac{\ddot{\phi}}{H \dot{\phi}} \, . 
\end{eqnarray}
The reason why we introduced $\epsilon_H$ is to keep the analysis fully general: in the backreaction dominated case~\footnote{Note, however, that in the backreaction case one might also worry that $\zeta$ could have a non-negligible contribution from gauge field fluctuations at horizon crossing, which is not taken into account by the above eq.~(\ref{PzetaT}).}, in fact, this parameter differs from the previous definition, $\epsilon=\dot{\phi^2}/(2H^2 M_P^2)$.

We can also compute the tensor-to-scalar ratio. Tensor perturbations ($h$) have couplings to gauge fields $\sqrt{\epsilon}$ smaller than the scalars and so are more difficult to thermalize. Therefore, while we assume here $\zeta$ to be thermal, we analyze first the most likely case of a standard vacuum spectrum~\footnote{We neglect here possible 1-loop contributions to $P_h$. In principle, in fact, there could also be a region where the tensors are non-thermal but have relevant 1-loop corrections to their 2-point function.}, for tensors.  In this case the tensor to scalar ratio is suppressed, because scalar perturbations are enhanced, giving:
\begin{eqnarray}
r \equiv \frac{P_h}{ P^\text{therm}_\zeta}  = 16 \, \epsilon \frac{H_*}{2 T_*} \, .
\end{eqnarray}
This case is phenomenologically very interesting because it would help polynomial large field models ($V(\phi)\propto \phi^n$, with $n>1$), but also others models of inflation where $\epsilon$ and $\eta$ are comparable, to agree with observations. In fact the observational bound $r\lesssim 0.1$ usually puts a quite stringent bound on $\epsilon$, while in our case if $T_*/H_*$ is sufficiently large the bound is relaxed and those models are not anymore in tension with observations.

We also analyze for completeness the case of thermalized tensor modes. Since they behave as massless scalar fields in de-Sitter, the result would be similar to eq.~(\ref{PzetaT}), {\it i.e.}, enhanced by $2T_*/H_*$ over the vacuum case
\begin{eqnarray} \label{thermal PS}
P_h^\text{thermal} = \frac{H_* T_*}{\pi^2 M_p^2} \,.
\end{eqnarray}
In this case the tensor-to-scalar ratio would remain unchanged, $r=16 \epsilon$, while the tensor tilt  would be non-trivial:
\begin{eqnarray} \label{tensor tilt}
n_T \equiv \frac{d \ln P^\text{therm}_h }{d \ln k} &=& -2 \epsilon_H + \frac{d \ln(T_*/H_*) }{d \ln k} \, .
\end{eqnarray}
 This regime would actually be very interesting. In fact, normally the tensor tilt is necessarily red ($n_T<0$), reflecting the fact that the Hubble constant decreases during inflation (and so $\epsilon_H>0$); here, instead, we could find a blue spectrum depending on the behavior of $T_*/H_*$. It is however difficult to have thermalization of tensors, due to their weaker coupling, as we will see in section~\ref{thermalmass}.

\subsection{Constraints on $f$ and $\xi$}

We start by briefly reviewing the main observational constraints on $\xi$ derived in the literature and then discuss how are they affected by thermalization.

If the inflaton perturbations are not thermal, and in absence of backreaction, several constraints on $\xi$ have been derived. Non-Gaussianity on CMB scales constrains $\xi \lesssim 2.5$ \cite{Barnaby2011, Bartolo2014, Ferreira2014a}. Moreover, perturbativity on the loop expansion for the cosmological correlators requires~\cite{Ferreira2015a}
\begin{eqnarray} \label{Perturbative Constraint}
\frac{H^2}{f^2} \frac{e^{2\pi \xi}}{16 \pi^2 l} <1 , 
\end{eqnarray}
where $l$ is some loop factor. For $l \simeq 10^{2}$ and imposing $P_\zeta^\text{vac}= 2.2 \times 10^{-9}$, the previous bound constrains $\xi \lesssim 3.5$ although specific 1-loop computations suggest a weaker bound $\xi \lesssim4.4$ \cite{Ferreira2015a, Peloso2016}. 

Another important threshold corresponds to the backreaction of gauge fields, in which the standard slow-roll regime driven by gravitational friction does not apply anymore. This happens if 
\begin{eqnarray} \label{backreaction1}
V'(\phi) \simeq 3H \dot{\phi} \ll \frac{\langle F \tilde{F} \rangle}{f} \, .
\end{eqnarray}
In the case without collisions we have $\langle F \tilde{F} \rangle \simeq  10^{-4} H^4 e^{2\pi \xi} /\xi^4$ \cite{Anber2009}, which means that backreaction is absent if $f/H\gg 4\times 10^{-3} e^{\pi \xi}/\xi^{5/2}$.

However, we stress here that these constraints do not directly apply if the system is thermalized simply because the occupation numbers completely change and new constraints should be derived. In fact, as we mentioned before, since thermalization moves particles from the horizon size to smaller scales, we expect the thermal case to be less constrained, as we discuss here.

But there are some simple constraints that we can directly apply. Namely, irrespectively of whether the spectrum of perturbations is thermal or not, the vacuum spectrum cannot exceed the observational bound, $P_\zeta^\text{vac}  \leq P_\zeta^\text{obs} \equiv 2.2 \times 10^{-9} $. The reason is that vacuum fluctuations are a lower bound and adiabatic perturbations cannot be erased once they cross the horizon~\cite{Linde2005}.
If we now recall the definition of $\xi$, this imposes a lower bound on $f$,
\begin{eqnarray}\label{ObsConstraint}
\frac{4\pi f \xi}{H}=  \frac{1}{\sqrt{P^\text{vac} _\zeta}}  \, \gtrsim \,2 \times 10^4 \qquad \Rightarrow \qquad \frac{f}{H} \gtrsim \frac{2 \times 10^3}{ \xi} \, .
\end{eqnarray}

\begin{figure} 
	\centering
	\includegraphics[scale=0.45]{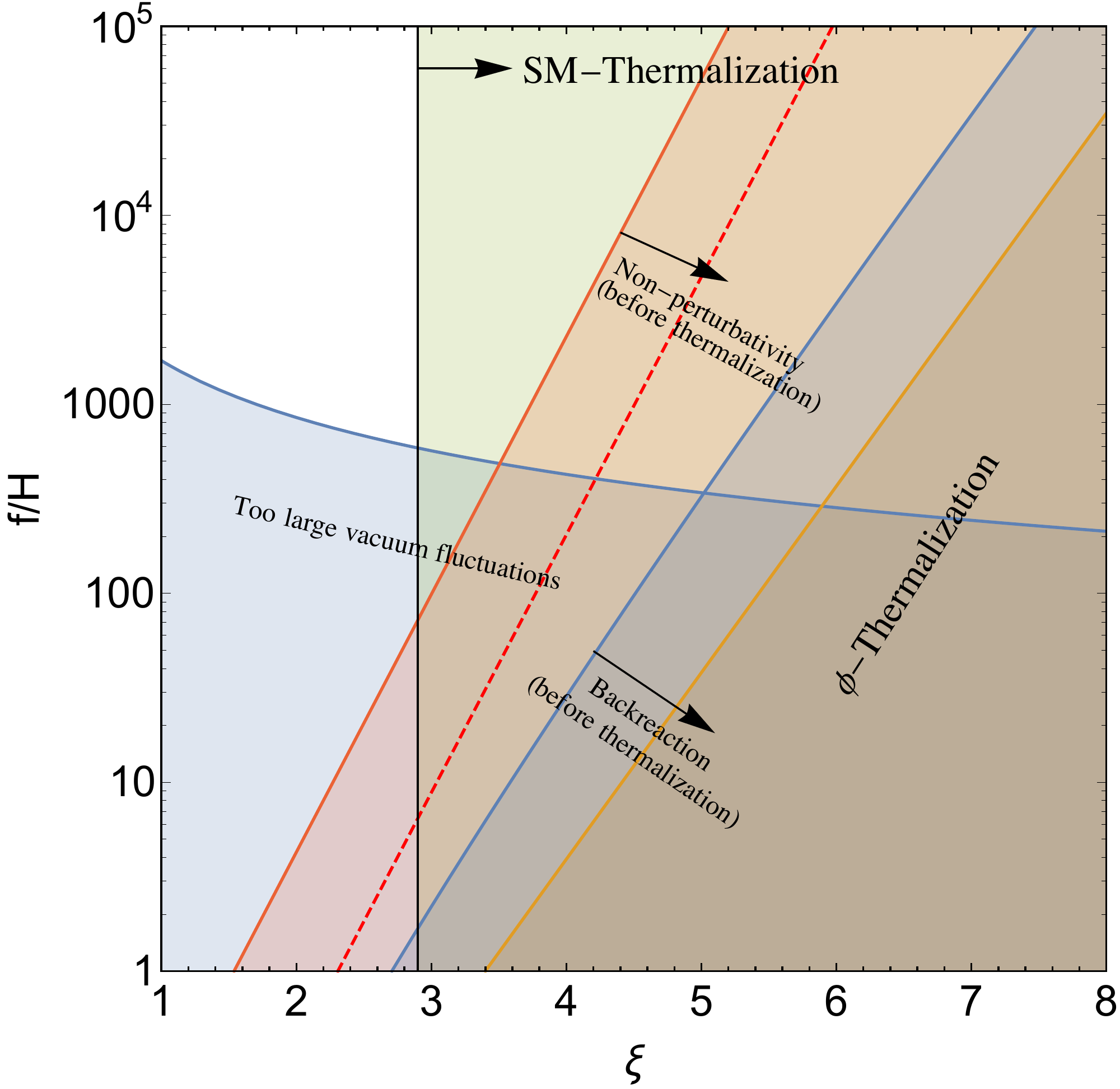}
	\caption{\label{constraint} Green: gauge fields thermalize through SM interactions. Red: Region where the perturbative expansion breaks down in the absence of thermalization (straight line derived from a parametric estimation and dashed from an explicit 1-loop calculation \cite{Ferreira2015a, Peloso2016}). Purple: backreaction region in the absence of thermalization. Orange: inflaton perturbations thermalize (and also gauge fields thermalize, even in absence of SM interactions). Blue: Vacuum fluctuations larger (or equal) than the observed value.}
\end{figure}

In fig.~\ref{constraint} we plot the different thresholds for thermalization derived in sections~\ref{estimates} and~\ref{numerics}. Jointly with the thermalized regions we overlap the perturbative constraint eq. (\ref{Perturbative Constraint}) (derived in absence of thermalization) for $l=10^{2}$, the backreaction threshold (also in absence of thermalization) eq.~(\ref{backreaction1}) and the observational constraint eq.~(\ref{ObsConstraint}). 
In absence of SM interactions the phenomenological allowed region for $\phi$ to be thermalized is such that one would need to take into account higher order loop corrections to the propagators as well as backreaction, at least before the onset of thermalization. Therefore, a dedicated study is required.
However, we stress that in the presence of SM interactions the system can thermalize before backreaction although, in this case, it is also crucial to take into account thermal masses to establish when $\phi$ thermalizes and when backreaction is relevant. We address this issue in the next section.

 Finally, we argue why, generically, the constraints on $\xi$ are very much alleviated when the system thermalizes. The reason is two-folded. First, the fact that energy moves from the horizon scale to the UV, with a consequent decrease in the particle number, changes the peak of the integrals from the horizon scale to the temperature scale. This will affect all correlators. Moreover, odd correlators are further suppressed by the following reason. The interaction Hamiltonian associated with the axial coupling is given by \cite{Barnaby2011, Ferreira2014a}
\begin{eqnarray}
H_\text{int} = \frac{\xi}{2} \int d^3 x \, \zeta \,F_{\mu \nu} \tilde{F}^{\mu \nu}. 
\end{eqnarray}
Therefore, all odd correlators will necessarily be proportional to, at least, one power of $\left< F  \tilde{F} \right> \propto d/d\tau (N_+ -N_-)$. In a thermal bath $N_+$ and $N_-$ tend to equilibrate through scatterings and so any parity asymmetry tends to be suppressed. This asymmetry would be completely erased if the two helicities had exactly the same dispersion relation. However this is not the case, since the axial coupling changes the dispersion relations of $\omega_{+}$ and $\omega_-$  in different ways, and so a small asymmetry should remain there, proportional to $\xi H$, even in thermal equilibrium. 

Computing loop corrections in the thermal regime requires a dedicated study~\cite{future}. Nonetheless, in appendix~\ref{NG} we provide a rough derivation of the parametric dependence of the non-Gaussian parameter $f_{NL} \simeq \left< \zeta^3 \right>/\left< \zeta^2 \right>^2$ with the temperature, considering the case where gauge fields are thermalized but $\phi$ is not. We find 
\begin{eqnarray}
f_{NL} \simeq c \, \xi^4  P_\zeta^\text{vac}  {\cal O}\left( \frac{T^4}{ H^4}\right)\, ,
\end{eqnarray}
where $c$ is a small number containing inverse powers of $(2\pi)$ and the result of the angular integration and so needs to be derived in a more accurate computation.
If we consider the case of instantaneous thermalization of the plasma, then, the energy density in the gauge fields is  simply  given by the initial energy $\rho_{\g} \simeq 10^{-4} H^4 e^{2\pi \xi}/\xi^3$ \cite{Anber2006}. Therefore, the plasma would have a temperature $\bar{T} \simeq \rho^{1/4} \simeq 0.1 H e^{\pi \xi/2}/\xi^{3/4}$ which means that $f_{NL}$ would be proportional to $e^{2\pi \xi}$. Comparing this result with the non-thermal case where $f_{NL} \simeq 10^{-7} e^{6\pi \xi} P_\zeta^\text{vac}/\xi^8$ \cite{Barnaby2011} there is a parametrical suppression of $e^{-4\pi \xi}$ which translates into a weaker constraint for $\xi$. In particular, assuming $10^{-4}> c>10^{-7}$ and using the constraint $f_{NL} < {\cal O}(10)$ \cite{Ade:2015ava} one would find  $\xi\lesssim 6-7$. However such a high temperature $\bar {T}$ is actually {\it not} what we finally expect, as we discuss in the next section. In reality, we expect a much smaller temperature to be reached, thus relaxing even more the constraints.
In appendix \ref{NG} we also argue that, if $\phi$ is also thermalized, $f_{NL}$ can be at most multiplied by an additional power of $T/H$
\begin{eqnarray}
f^\text{thermal}_{NL} \lesssim d \, \xi^4  P_\zeta^\text{vac}  {\cal O}\left( \frac{T^5}{ H^5}\right)\, ,
\end{eqnarray}
where $d$ is another small coefficient.

\section{Effect of thermal masses \label{discussion}} \label{thermalmass}

So far we have studied the conditions for thermalization to occur due to the initially large occupation number of gauge bosons.
However, it is more subtle to understand what happens after thermalization is established.

As we discussed in previous sections, due to the presence of an imaginary dispersion relation for $\g_+$, it is unclear how is the shape of the distribution at low momenta, $k/a<2 \xi H$.  
In fact such modes are exponentially produced, although in a finite window of time, and so one expects relevant deviations from a BE distribution such as the presence of a peak at low momentum. The $\gamma_-$ polarization, instead, should be correctly described by a BE distribution with a dispersion relation $\omega^2=k^2+m^2$, where $m^2=-2 k \xi/ \tau$, while $\delta\phi$, if thermalized, is described by a massless BE distribution. 

However a very important ingredient modifies the above discussion: when thermalization happens, gauge fields typically\footnote{This certainly happens in the well-known case of having interactions with charged SM particles. It could also happen for the case of only $\phi$-mediated interactions, but this would require performing the one-loop thermal correction, which we postpone to a subsequent publication~\cite{future}.} develop a thermal mass $m_T \propto g T$, where $g$ is the gauge coupling, as a result of having a non zero density of particles correcting the propagator at 1-loop. Note, importantly, that $\phi$ instead does not get a thermal mass, since its interactions only renormalize its kinetic term~\cite{future}. We leave to the future a more careful study of the effect of thermal masses, but we anticipate here some interesting features.

Let us parameterize the thermal mass as
\begin{eqnarray}
m_T\equiv \bar{g} T \, .
\end{eqnarray}
In the case of SM fermions $\bar{g}^2=g^2 \sum_i q^2_i/6 $, where $q_i$ is the charge of a given fermion. At energies of around $10^{14}$ GeV this gives $\bar{g}^2 \simeq 0.3$.
If the thermal mass is included in the equation of motion for the gauge fields it will compete with the instability due to the axial coupling, giving
\begin{eqnarray} \label{EOM for A with mass}
A_\pm'' + \omega^2_T(k) A_\pm =0,  \qquad \omega_T(k)=\left(k^2 \pm \frac{2k \xi}{\tau}  + \frac{m_T^2}{H^2 \tau^2} \right) \, .
\end{eqnarray}
When $m_T\geq \xi H$ the mass completely shields the instability band. Therefore, gauge fields cannot be produced  and we might expect an {\it equilibrium} temperature \begin{eqnarray} T_{eq}= \frac{\xi H}{ \bar{g}} \, .\end{eqnarray}
In fact, starting from a non-thermal case, when we reach thermalization the plasma has an initial temperature, $\bar {T}$, typically higher than $T_{eq}$. Thus, at that point the particle production stops and the plasma of particles is expected simply to redshift as radiation, as $a^{-4}$, which is equivalent to say that temperature decreases as $1/a$. When the temperature drops below $T_{eq}$ an instability band reappears, most likely as a narrow band centered at $|k\tau|\approx \xi H$, which is the minimum of $\omega_T(k)^2$. At this point we expect the system either to reach an almost stationary state slightly below $T_{eq}$ or perhaps an oscillatory behavior around such temperature.

When the temperature drops, thermalization could be less efficient. We can estimate the condition for the plasma to remain in thermodynamic equilibrium by comparing $\sigma_{eq} \cdot n_{eq}$ with $H$, using the equilibrium number density $n_{eq} \simeq T^3$ and the typical cross section $ \sigma_{eq} \simeq \bar{\alpha}^2/T^2$, where $\bar{\alpha}^2\equiv g^4/(4\pi)^2 \sum_i q^4_i$ is summed over different species. In the SM case at $10^{14}$ GeV, $\bar{\alpha}^2 \simeq 0.05$. As a result we get:
\begin{eqnarray}
\frac{T}{H} \gg \frac{1}{\bar{\alpha}^2}  \simeq \, 20 \, .
\end{eqnarray}
To estimate if the plasma keeps thermalized down to $T_{eq}$ we should fulfill the condition  
\begin{eqnarray} \label{SMtherm2}
\frac{T_{eq}}{H} \gg \frac{1}{\bar{\alpha}^2} \qquad \Rightarrow \qquad \xi \gg \frac{\bar{g}}{\bar{\alpha}^2}  \simeq 10 \, .
\end{eqnarray}
For a given $g$ and a given set of fermions, this condition can indeed be met for sufficiently large $\xi$ while when considering only SM fermions we get a more precise condition on $\xi$. Note however that, even if the condition is not met in the range $\bar{T}>T>T_{eq}$ we still expect the plasma simply to redshift down to $T_{eq}$, because the instability remains screened by the thermal mass.

\subsection*{Equilibrium temperature} 

In the rest of this section we will discuss the case where the system reaches a stationary temperature $T_{eq}$. In that case we expect a narrow band of the instability around $|k\tau|\approx \xi H$ to be constantly present, generating a peak in the $\gamma_+$ distribution, whereas we expect a thermal distribution away from such narrow peak, where $\omega_T(k)^2>0$. Observing this dynamics should be numerically feasible but we postpone further investigation of this important issue to future work. We now briefly discuss possible interesting observational consequences of such cases.

We have to distinguish between the case with SM interactions and the case with only $\phi$-mediated interactions. In the first case, as we saw before, we can reach a thermal state if $\xi\gtrsim 2.9$. Then we need to check if $\phi$ thermalization is efficient by using, as above, the thermal condition $\sigma_{eq} \cdot n_{eq} \gg H$, with $ \sigma_{eq} \approx T^2/f^4$,    so that:
\begin{eqnarray}
T\gg (f^4 H)^{1/5} \, ,
\end{eqnarray}
which should be true at $T=T_{eq}$, thus imposing
\begin{eqnarray} \label{phitherm2}
\xi \gg \bar{g} \left(\frac{f}{H} \right)^{4/5} \, .
\end{eqnarray}
If the above condition is met we can use the results from the previous section to make predictions for the spectrum of perturbations. 

If $\phi$ is thermal from eq. (\ref{PzetaT}) we have that:
\begin{eqnarray} \label{spectherm2}
P^\text{therm}_\zeta = \frac{\xi}{\bar{g}} \frac{H^4_*}{2 \pi^2 \dot{\phi}_*^2 } = \frac{H^2}{8 \pi ^2 f^2 \bar{g} \xi } \, ,
\end{eqnarray}
and the spectral index becomes simply 
\begin{eqnarray}
n_s-1\equiv \frac{d \ln P^\text{therm}_\zeta }{d \ln k} = -6 \epsilon_H+2 \eta + \frac{\dot{\xi} }{H \xi} =  -4 \epsilon_H+\eta  \, .
\end{eqnarray}
We remind the reader that we are still working under the assumptions listed at the beginning of section~\ref{phenomenology}, which are valid in the non-backreacting regime and imply that $\zeta$ is conserved on superhorizon scales.

\begin{figure} 
	\centering
	\includegraphics[scale=0.45]{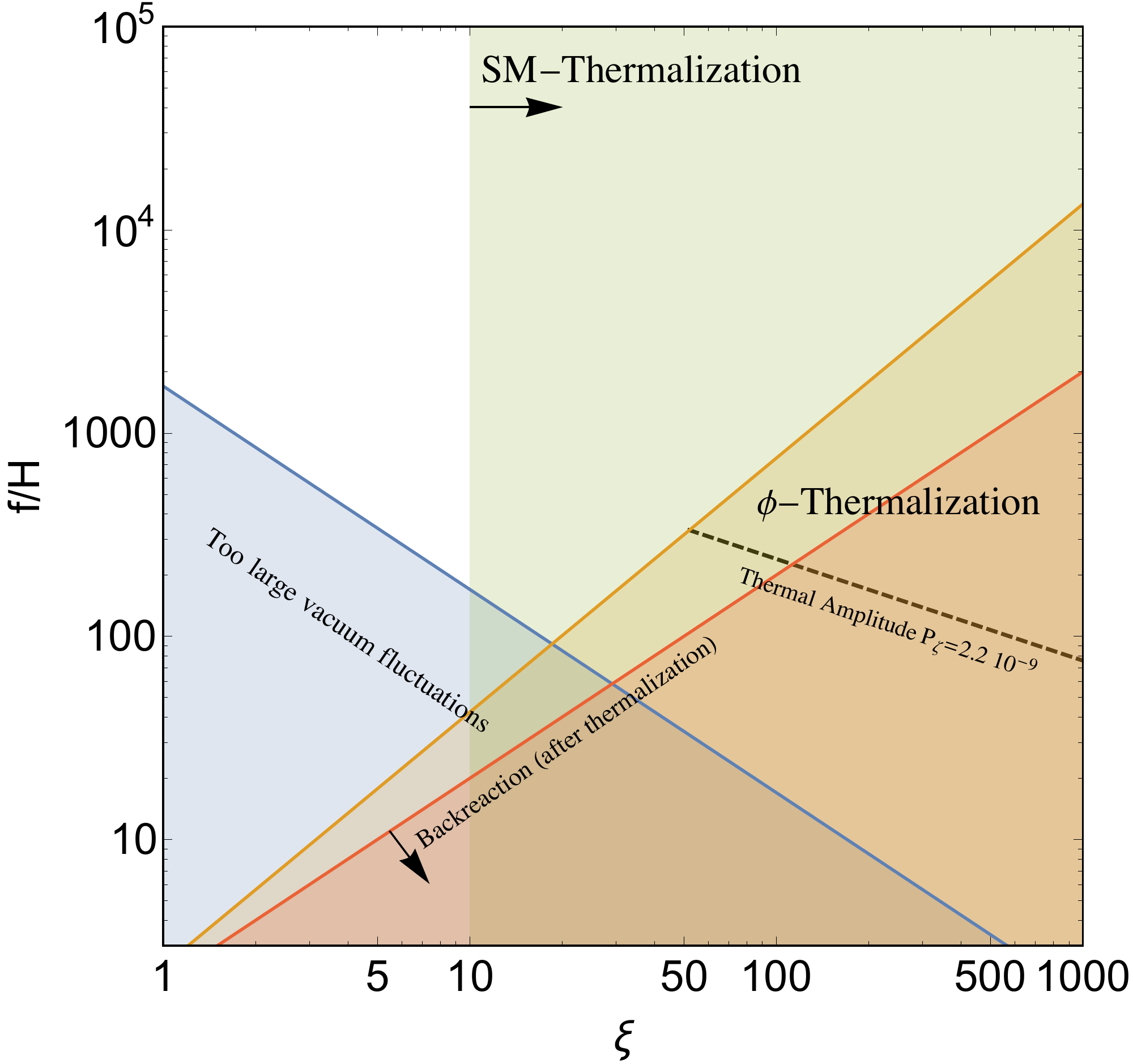}
	\caption{\label{constraint2} Same conditions as fig.~(\ref{constraint}) but in the presence of a thermal mass and assuming an equilibrium temperature of $T_{eq}=\xi/\bar{g}$. In this plot we fix $\bar{g}=0.5$. We also show as a dashed line the condition that a thermal spectrum fits the observational value for $P_\zeta$.}
\end{figure}

In the case of tensor modes the couplings are gravitational, suppressed by $1/M_p$, and so the cross section would be given by $\sigma_{eq} \approx T^2/M_p^4$ and so in order to reach thermalization one would need $T \gg (H M_p^4)^{1/5}$. This requires a temperature larger than the inflationary energy, $\rho_{inf}^{1/4}=3H^2 M_p^2$, and so, for that reason, it is not possible to thermalize the tensor modes during inflation at this equilibrium temperature. Therefore, assuming tensor modes to be in the vacuum the tensor to scalar ratio would be given by
\begin{eqnarray}
r \equiv \frac{P_h}{ P^\text{therm}_\zeta}  = 16 \, \epsilon \frac{H}{2T} \approx 8\epsilon \frac{\bar{g}}{\xi} \, .
\end{eqnarray}
As we mention before, thermalization of $\phi$ leads to a suppression of the tensor to scalar ratio which, at the equilibrium temperature, would be $\bar{g}/(2\xi)$. If we look at fig.~\ref{constraint2}, this amounts to at least an ${\cal O}(10^{-2})$ suppression.  

We can also try to estimate whether there is relevant backreaction on the scalar field equation of motion, as follows. The energy extraction from $\phi$ is given by $\langle F \tilde{F} \rangle \dot{\phi}/f$ which we can estimate by using conservation of energy density in the gauge fields $\rho_\gamma$:
\begin{eqnarray}
\dot{\rho}_\gamma+4 H \rho_\gamma= \frac{\dot{\phi}}{4f} \langle F \tilde{F} \rangle \, .
\end{eqnarray}
In a stationary situation $\dot{\rho}_\gamma\approx 0$ while backreaction is achieved if $\langle F \tilde{F} \rangle/(4f)\gtrsim 3 H \dot{\phi}$. These two conditions then require
\begin{eqnarray} \label{backtherm2}
\xi \gtrsim 3 \,\bar{g}^2 \frac{f}{H} \, .
\end{eqnarray}

Finally, using the estimation of non-Gaussianity in the case where $\phi$ is thermal, eq.~(\ref{NGresult2}), we can also find what would be a conservative viable region for $T_{eq}= \xi/\bar{g}$ by requiring the non-Gaussian parameter $f_{NL} < {\cal O}(10)$. This would impose $\xi \lesssim 20-50$ which corresponds to a small viable window in fig. (\ref{constraint2}).

In fig.~\ref{constraint2} we plot the different constraints and phenomenological windows as in fig.~\ref{constraint}, but now including the effects of a thermal mass and assuming a stationary regime. The different regimes are described by eqs. (\ref{SMtherm2}), (\ref{phitherm2}), (\ref{backtherm2}), (\ref{spectherm2}) and (\ref{ObsConstraint}). Note that, contrary to fig.~\ref{constraint}, the region where $\phi$ is thermalized is not completely inside the backreaction region and so there is some parameter space where $\phi$ can thermalize without backreaction.

In the case of oscillatory temperatures around $T_{eq}$ the above description remains probably roughly correct although in that case we should expect larger deviations from the thermal predictions as well as superimposed oscillations in the relevant quantities at horizon crossing.  Clearly, if such oscillations are large the model is observationally ruled out. However, interestingly, if oscillations are within the observational bound but non-negligible, this would result in oscillatory power spectra, which could be another striking evidence of this mechanism.

%%%%
\section{Conclusions}  \label{conclusions}

In this work we have analyzed the system composed of an inflaton ($\phi$) coupled to a gauge field ($A_\mu$) through a CP-odd term with strength $1/f$. One helicity of the gauge field, $\g_+$, is known to develop an instability with exponential particle production at a rate controlled by the parameter $\xi=\dot{\phi}/(2f H)$, for modes with momenta in the range $(8\xi)^{-1} H \lesssim k/a \lesssim 2\xi H$. We have included in this scenario the effect of scatterings through a set of Boltzmann-like equations. The scattering rates are Bose-enhanced by the large occupation numbers of $\g_+$ and can drive the system to thermal equilibrium, thus leading to a setup that we dubbed {\it thermalized axion inflation}. We have analyzed two situations: (1) only scalar-gauge field interactions, as in the case of a ``dark photon"; (2) identifying the gauge field with a Standard Model gauge boson (abelian or non-abelian). We stress that (2) is probably the best motivated case, since this system anyway needs to be coupled to the SM to provide reheating, and it turns out to be the one in which we have good control of the theory, since thermalization happens before backreaction and before perturbativity constraints.

In the first case the most relevant scatterings are $\g_+ \g_+ \rightarrow \g_+ \g_+$. By solving the associated Boltzmann-like set of equations of the system we find that $N_{\g_+}$, $N_{\g_-}$ and $N_{\phi}$ evolve to a Bose-Einstein distribution when $\xi \gtrsim 0.44 \ln(f/H) +3.4$. Note, however, that due to the axial coupling the gauge fields have a modified dispersion relation, imaginary in the case of $\g_+$, which is expected to distort the distribution for low momenta $k/a<2\xi H$. In a phenomenologically realistic model one has to require the power spectrum of curvature perturbations to be compatible with observations, and that imposes a lower bound on $f/H$, which then implies thermalization at $\xi\gtrsim 6$. At such values, however, we would have to assume that the occupation numbers are not significantly altered by loop effects~\cite{Ferreira2015a}, which are important already at $\xi \gtrsim 3.5$. Moreover, for such a large $\xi$ the gauge fields would also backreact on $\phi$ and on $\delta\phi$, so a dedicated study is needed. We will return to this point in a future work.

In the second case, where we consider SM interactions with the gauge field, thermalization is easier to achieve because the cross sections are not suppressed by powers of $H/f$. This case is also more predictive because $\xi$ is the only free parameter. For example, if the gauge field is the one associated to $U(1)$ hypercharge and considering only particle anti-particle productions, we find that thermalization happens already at $\xi \gtrsim 2.9$. A similar condition on $\xi$ applies if instead we consider the gauge field to be a gluon and include self-interactions. This is of great interest because in this case one can have a thermal bath of particles well under control, before perturbativity constraints become relevant and also before backreaction of gauge fields on $\phi$ becomes important.

In both cases thermalization has profound implications for the phenomenology of axion inflation. To elaborate on this we have further pointed out that, after thermalization is reached, the system should evolve in a very different regime, due to the presence of thermal masses $m_T=\bar{g} T$ for the gauge bosons, which tends to screen the instability of $\gamma_+$. As a result we expect either an oscillatory behavior or a stationary solution at a temperature given by $T_{eq}=\xi H/\bar{g}$. Note that this regime is now only linearly dependent on $\xi$ and not exponential anymore and so all constraints on $\xi$ should become much weaker and should be properly readdressed. We described such behavior only qualitatively in this work but we plan to address these points in full detail in the future.

Assuming the system reaches a stationary solution, or one where the temperature changes adiabatically, we derived very interesting consequences for cosmological observables such as the scalar and tensor perturbations, under the assumption that $\phi$ is thermalized, which can happen in the region shown in fig.~\ref{constraint2}.  In this regime one can also find a region of parameters where $\phi$ can thermalize without backreacting on the background. Note also that in this case the power spectrum of curvature perturbations can be obtained simply by setting its value at horizon crossing to be that of a thermal state, assuming it will be conserved on superhorizon scales.
This leads to a result for $P_\zeta$ that is enhanced compared to the vacuum case by the ratio of particle numbers between the thermal state, $T_*/H_*$, and the vacuum, $1/2$. We have also computed the tilt of the power spectrum to be $n_s-1=4\epsilon_H-\eta$, which differs from the standard slow-roll formula because of  the time dependence of $T_{eq \, *}/H_*$.

Regarding the tensor modes, because of their Planck suppressed couplings it does not seem possible for them to thermalize, at least at the equilibrium temperature. For that reason we expect the tensors to remain in the vacuum and so, in the case in which $\phi$ is thermalized, a tensor to scalar ratio suppressed by $H_*/(2T_*)$ with respect to the standard formula. This is very interesting because it can allow for a reconciliation between models of inflation where $\epsilon$ and $\eta$ are comparable, {\it e.g.} polynomial large field models, with observations.

Finally, although we did not present here a detailed study of non-Gaussianities, we provided strong arguments, and one particular estimation, which justify why the previous constraints do not apply to our case and why, generically, we expect them to be less restrictive in the thermal regime. 
Thermalization redistributes in fact the occupation numbers: it depletes the highly populated horizon-sized modes and it populates higher $k$ modes. As we have argued, loop corrections to cosmological correlators will then be generically smaller than in the non-thermal case. In addition, odd correlators of $\zeta$ are further suppressed, since thermodynamic equilibrium tends to restore parity, driving $\langle F\tilde{F} \rangle$ to small values. We estimated the non-Gaussian parameter $f_{NL} $ to be proportional to\footnote{This estimation holds if $\phi$ is not thermalized, while in the opposite case we showed that there can be, at most, one extra power of $T/H$.} $ P_\zeta^\text{vac} \,T^4/H^4$, which corresponds to a large parametric suppression of $e^{-4\pi \xi}$ compared to the non-thermal case.  This seems promising since it could allow for interesting regimes with large $\xi$, such as the $\phi$-thermalized regime and the backreacting regime~\cite{future}, to be in agreement with observations.

To sum up we provided a working model in which {\it during} inflation a thermal bath with a possibly large temperature is present, leading to new implications in model building and to new observational features. One of the crucial ingredients of this setup is the fact that the axial coupling respects a continuous shift symmetry and so cannot induce thermal mass corrections. This is true in the U(1) case, while in non-Abelian theories it can be broken by non-perturbative effects. If the thermalized backreacting regime can be achieved in full control it could even remove the standard need of an inflaton flat potential, by the help of the dissipative friction. Moreover reheating is already incorporated in this scenario and obtained just as a transition when the thermal bath starts dominating over the potential energy of the inflaton.  

${}$\linebreak
\emph{\textbf{Acknowledgments:}}
We thank K.~Tywoniuk, J.~Garriga, C.~Germani, F.~Mescia, E.~Verdaguer, K.~Tobioka and M.~Sloth for comment and useful discussions. This work is supported by the grants EC FPA2010-20807-C02-02, AGAUR 2009-SGR-168, ERC Starting Grant HoloLHC-306605 and by the Spanish MINECO under MDM-2014-0369 of ICCUB (Unidad de Excelencia “Maria de Maeztu”).

\appendix

\section{Appendix} \label{appendix}

\subsection{Fourier conventions}

We quantize scalars ($\delta\phi$ or $\zeta$) and gauge fields as
\begin{eqnarray} \label{Fourier transform}
\zeta(x,\tau ) &= &\int \frac{d^3k}{(2\pi)^3} e^{-ik \tau} \left[\zeta_k a_k + \zeta_k^* a_{-k}^\dagger \right] \nn \, ,\\
\vec{A} (x,t) &= &\int \frac{d^3k}{(2\pi)^3} e^{-ik \tau} \sum_{\si} \vec{e}_{k,\si} \left[A_{k,\si} b_{k,\si}+ A_{k,\si}^* b_{-k,\sigma}^\dagger \right]  \, ,
\end{eqnarray}
where we have used the Coulomb gauge ($A_0=0$) and where the creation and annihilation operators satisfy
\begin{eqnarray}
\left[a_k, a_{-q}^\dagger \right] &=&(2\pi)^3 \delta^{(3)}(k+q) \, , \\
\left[b_{k,\si}, b_{-q,\si'}^\dagger \right] &=& (2\pi)^3 \delta_{\si,\si'} \delta^{(3)}(k+q) \, .
\end{eqnarray}
The polarization vectors satisfy the following identities
\begin{eqnarray} \label{pol ident}
\vec{k }\times \vec{e}_{\pm} (\vec{k}) = \mp \vec{k} \vec{e}_{\pm} (\vec{k}), \qquad \vec{e}_{\si} (\vec{k}) \cdot \vec{e}_{\si'} (\vec{k})^* = \delta_{\si \si'}, \qquad e_{\pm}(\vec{k})= e_{\mp}(-\vec{k})=e_{\mp}(\vec{k})^*  \, .
\end{eqnarray}

\subsection{Decay estimation from 1-loop correction \label{decays}}

In section~\ref{decays estimate} we presented the 1-loop correction to the two point function which we then used as an estimation of the decay rate.
In this appendix we present some of the intermediate steps which took us to eq. (\ref{e2}).
Starting from eq. (\ref{start}) by taking the time derivative of the two point function we arrive at
\begin{eqnarray}
\frac{d}{d \tau}\left< u_k u_k \right>_\text{loop}  (\tau)= &&2H a^2   \left< \delta \phi_k \delta \phi_k \right>_\text{loop}   (\tau)+a^2 \int_{-\infty}^{\tau}  \frac{d \tau''}{(2\pi)^{12} (2f)^2}\int d^3 q \left| \vec{e}_+ (\vec{q}) \cdot \vec{e}_-(|\vec{k}-\vec{q}|) \right|^2\times  \nn \\
&& \times  \left< \left[A'_q (\tau) A_{|\vec{k}-\vec{q}|}(\tau) |\vec{k}-\vec{q}| \delta \phi_k(\tau) ,\left[A'_{|\vec{k}-\vec{q}|}(\tau'') A_q(\tau'')  q \delta \phi_k(\tau'') , \delta \phi_k (\tau)\delta \phi_k(\tau) \right]\right] \right> + \nn \\ && + \, \text{perm.} 
\end{eqnarray}	
The time integrations are dominated by the latest time simply because, before thermalization, that is when the gauge field particle number has its largest value. This is only true until horizon crossing, afterwards the integral goes quickly to zero.
Therefore, we simplify the previous expression by replacing the  time integration by its integrand evaluated at $\tau$ times the integration interval which we approximate to be $\Delta \tau= 2 \xi/\text{min}(q,|\vec{k}-\vec{q}|)$ corresponding to the time at which the largest mode enters the resonant band and so where the process becomes non-zero. Thus, we arrive at
\begin{eqnarray}
\frac{d}{d \tau}\left< u_k u_k \right>_\text{loop}  (\tau)= &&2H a^2   \left< \delta \phi_k \delta \phi_k \right>_\text{loop}   (\tau)+ \frac{2 \xi a^2}{ (2\pi)^{12} (2f)^2}\int \frac{d^3 q}{\text{min}(q,|\vec{k}-\vec{q}|)} \left| \vec{e}_+ (q) \cdot \vec{e}_-(|\vec{k}-\vec{q}|) \right|^2\times  \nn \\
&& \times  \left< \left[A'_q (\tau) A_{|\vec{k}-\vec{q}|}(\tau) |\vec{k}-\vec{q}| \delta \phi_k(\tau) ,\left[A'_{|\vec{k}-\vec{q}|}(\tau) A_q(\tau)  q \delta \phi_k(\tau) , \delta \phi_k (\tau)\delta \phi_k(\tau) \right]\right] \right> + \nn \\ && + \,\text{perm.} 
\end{eqnarray}
Our goal is to rewrite this expression as powers of 2-point functions which we then associate with particle numbers. However, by isotropy the particle number should not depend on the angle, so it is convenient to perform the angular integral. At this level we do not have access to the generic mode functions in the presence of scatterings, so we parameterize this unknown by $b$ which was found in similar 1-loop computations to be ${\cal O}(10^{-3})$.
The drawback is, however, that we now need to specify the angle between $k$ and $q$ at which the angular integral would peak in order to compute $|\vec{k}-\vec{q}|$. We assume $|\vec{k}-\vec{q}| \simeq k + q$. Under these simplifications we arrive at
\begin{eqnarray}
\frac{d}{d \tau}\left< u_k u_k \right>_\text{loop}  (\tau)= &&2H a^2   \left< \delta \phi_k \delta \phi_k \right>_\text{loop}   + \frac{2 \xi a^2 b \delta^{3}(0)}{(2\pi)^{12} (2f)^2 }\int  \frac{dq q^3 |\vec{k}-\vec{q}|}{\text{min}(q,|\vec{k}-\vec{q}|)}   \times \nn \\
&& \times \left< \left[A'_q  A_{k+q} |\vec{k}-\vec{q}| \delta \phi_k,\left[A'_{k+q}A_q q \delta \phi_k , \delta \phi_k \delta \phi_k\right]\right] \right> + \text{perm.} \, ,
\end{eqnarray}
where we omitted the time argument in the right hand side because all quantities are evaluated at $\tau$. 
By a similar reasoning the first term in the right hand side of the previous equation gives
\begin{eqnarray} \label{e1}
\frac{4 H b \xi^2 a^2 \delta^{3}(0)}{(2\pi)^{12}(2f)^2 } \int  \frac{dq \, q^3 |\vec{k}-\vec{q}|}{\text{min}(q,|\vec{k}-\vec{q}|)^2 }  \left< \left[A'_q  A_{k+q} |\vec{k}-\vec{q}| \delta \phi_k,\left[A'_{k+q}A_q q \delta \phi_k , \delta \phi_k \delta \phi_k\right]\right] \right> \, . 
\end{eqnarray}
We only deal with subhorizon modes, thus, the first term is smaller or at most equal to the second term so we neglect it. Now, in order to write the expression as a particle number we approximate $|A'_k| \simeq k A_k$ and make use of the approximate relation $1/2 + N_u(k) \approx k |u_k|^2$. The commutator structure implies the expectation value to give an expression of the form
\begin{eqnarray}
\text{Im}\left[ \left<\delta\phi_k \delta\phi_k \right> \right]  \text{Im}\left[ \left<\delta\phi_k \delta\phi_k \right> \left<A_q A_q \right> \left<A_p A_p\right> \right]
\, . 
\end{eqnarray}
The imaginary part of the two point function can be roughly seen as the vacuum contribution, $(2\pi)^3$, while the real part would be proportional to $(2\pi)^3 (2N)$. This identification is accurate for thermal propagators and in fact by making this identity we would find that the previous equation has indeed a structure of particle numbers similar to those appearing in a standard decay expression which would be proportional to
\begin{eqnarray} \label{e2A} N_u(k) (1+N_{\g_+}(q)) (1+N_{\g_+}(k+q))- N_{\g_+}(q) N_{\g_+}(k+q) (1+N_u(k)) \, .  \nn
\end{eqnarray}
However, the 1-loop correction contains more processes than just the decay and for that reason we do not expect to arrive at that precise form. Nevertheless we can isolate the terms which do have the form of a decay and so we finally end up with
\begin{eqnarray} \label{e3}
&& \frac{dN_u(k)}{d\tau}=   -\frac{2 \xi \times 4 b}{(2\pi)^3 (2f)^2 a^2 k}  \times \nn \\
&& \times \int \frac{dq \, q^3 (k+q)}{\text{min}(q,k+q)} N_u(k) (1+N_{\g_+}(q)) (1+N_{\g_+}(k+q))- N_{\g_+}(q) N_{\g_+}(k+q) (1+N_u(k)) \, .
\end{eqnarray}
The extra factor of $4$ comes from identifying $\text{Re}[\left<X X\right> ] \propto 2 N_X$.

%%%%
%%%
\subsection{Matrix elements for scatterings \label{MatrixElements}}

In this appendix we provide the matrix elements for the scatterings depicted in fig.~\ref{diagrams}. 
We start by the $\gamma_+ \g_{+} \rightarrow \gamma_+ \g_{+}$ scattering which only has non-zero contributions from the s-channel
\begin{eqnarray} \label{++++}
M_{\g_+ \g_+ \g_+ \g_+}&=&  \frac{1}{4 a^2 f^2} \frac{\epsilon^{abcd} k^1_a k^2_b e_c(k^1) e_d(k^2) \epsilon^{w x y z} k^3_w k^4_x e^*_y(k^3) e^*_z(k^4) }{2 \left| k^\mu_1 - k^\mu_3 \right|^2} + 4 \,\text{perm.} \, ,
\end{eqnarray}
where $\epsilon^{abcd}$ is the 4-dimensional Levi-Civita tensor and $k^\mu$ are the 4-momenta of the particles involved in the scatterings. The processes involving $\gamma_+$ and $\gamma_-$ can be readily obtained from the previous result by making use of the identities in eq. (\ref{pol ident}) and summing over all the channels $u,s, t$. On the other hand the gauge field-axion scattering, $\gamma_+ \phi \rightarrow \gamma_+ \phi$, has instead s and t-channel contributions and is given by
\begin{eqnarray}
M_{ \g_+ \phi \g_+ \phi}&=&  \frac{1}{4 a^2 f^2} \frac{\epsilon^{abcd} k^1_a k^2_b e_c(k^1) \epsilon^{w x y z} k^3_w k^4_x e^*_y(k^3)  \eta_{dz}}{2 \left| k^\mu_1 - k^\mu_3 \right|^2} + k_2 \leftrightarrow k_3\, ,
\end{eqnarray}
where $\eta_{dz}$ is the Minkowski metric. Finally gauge field-axion conversion, $\gamma_+ \gamma_+ \rightarrow  \phi  \phi$, has matrix elements
\begin{eqnarray}
M_{ \g_+ \g_+ \phi \phi}&=&  \frac{1}{4 a^2 f^2} \frac{\epsilon^{abcd} k^1_a k^2_b e_c(k^1)  \epsilon^{w x y z} k^3_w k^4_x e^*_y(k^3)  \eta_{dz}}{2 \left| k^\mu_1 - k^\mu_3 \right|^2} + 2 \, \text{perm.}
\end{eqnarray}

\subsection{Particle number and Bogolyubov coefficients \label{bog}}

In curved space the particle number is typically defined as $N_k = |\beta_k|^2$, where \cite{Mukhanov:2007zz}
\begin{eqnarray}
\beta_k =  \frac{f_1(k) f'_2(k)-f'_1(k) f_2(k)}{W_k} \, ,
\end{eqnarray}
is the Bogolyubov coefficient which relates the mode functions between two different basis of states and $W_k= (f'(k) f^*(k)-f'^*(k) f(k))/(2i) $ is the Wronskian. If we want to know the particle number compared to the vacuum state we fix $f_1= e^{-ik\tau}/\sqrt{2k}$ and so
\begin{eqnarray}
N_k = |\beta_k|^2 = \frac{|f_2'(k)|^2+k^2 |f_2(k)|^2}{2k} \, ,
\end{eqnarray}
in agreement with our definition in eq. (\ref{number}).

\subsection{Numerical distribution of particle number \label{numbdist}}

\begin{figure} 
	\centering 
	\includegraphics[scale=0.35]{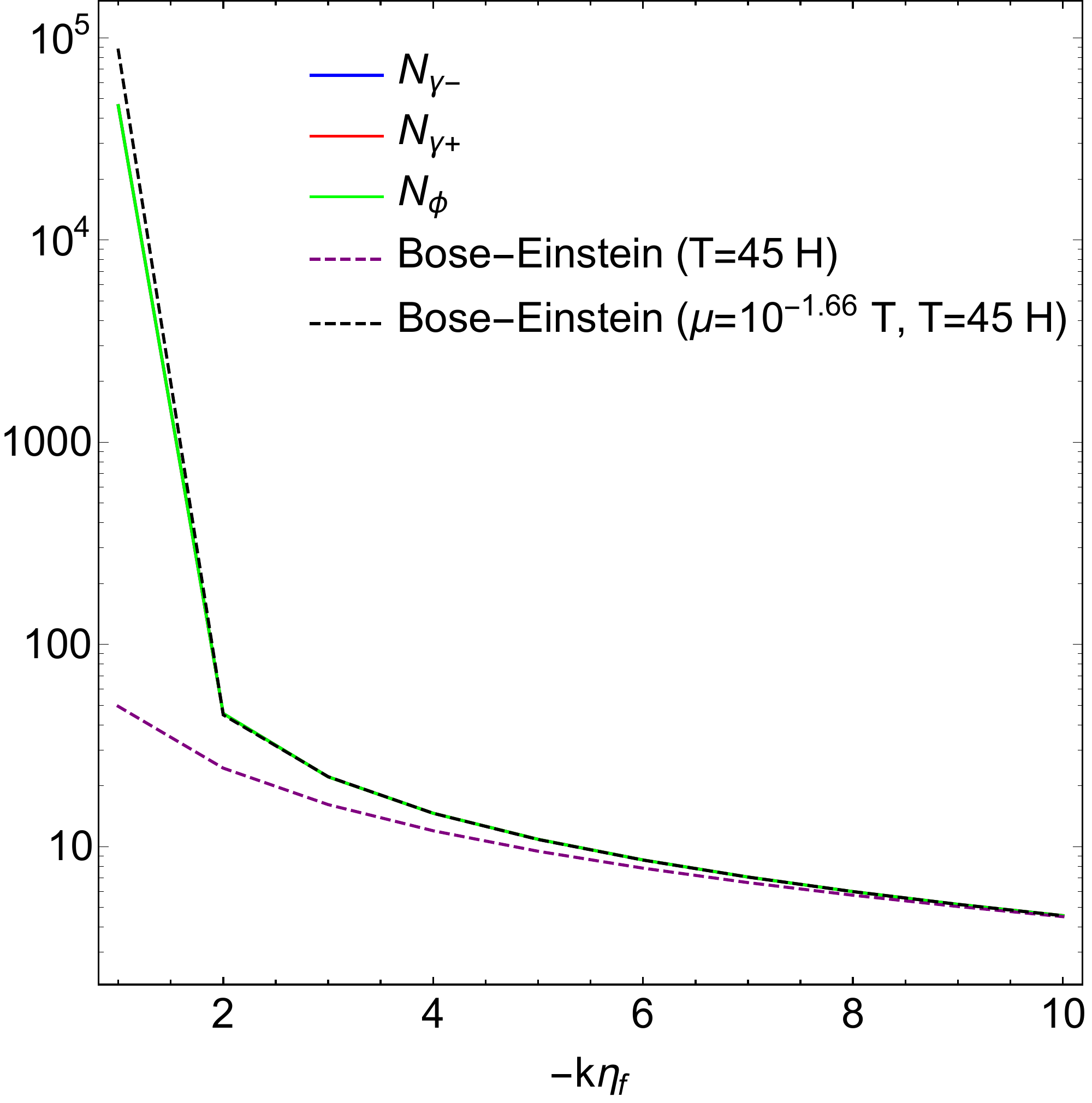}
	\caption{\label{dist} Particle number as a function of momenta at a time $\tau_f$ for $\gamma_{+}, \g_-$ and $\phi$. The dashed lines are fits using thermal distributions with and without chemical potential}
\end{figure}

In fig.~\ref{dist} we show the numerical distribution of particle number for the different species as a function of momenta and at the end of the simulation, $-k_\text{min} \tau=1$, for $\xi=3.9$ and $f=H$. In the same plot we show Bose-Einstein distributions fro massless particles with and without a chemical potential. We can see that, apart from the longest mode, a Bose-Einstein distribution with $T=45H$ is in rough agreement with the distribution.  The inclusion of a chemical potential $\mu = 10^{-1.66}T$ gives a perfect fit, including also the first mode.

\subsection{Estimation of loop corrections \label{NG}}

In the main text we have argued that in the thermal regime all constraints, such as non-Gaussianities, non-perturbative constraints and backreaction, are strongly modified by the fact that the energy moves from the horizon scale to the UV. Moreover, odd correlators are also further suppressed by parity arguments. 
In this appendix we give a rough estimation of the 1-loop correction to the 3 point function. We study the case where the photons are thermalized but the inflaton perturbations are not. We also derive an upper bound for the case with thermalized inflaton perturbations. This estimation is also easily generalizable to higher order correlators.
Let us start by the 3-point function. The 1-loop correction is proportional to
\begin{eqnarray}
\left< \zeta_k \zeta_q \zeta_p \right>_\text{1-loop} (\tau)= i^3 \int_{-\infty}^{\tau} d \tau_1 \int_{-\infty}^{\tau_1} d \tau_2 \int_{-\infty}^{\tau_2} d \tau_3 \left< \left[H_\text{int} (\tau_1) \left[H_\text{int} (\tau_2), \left[H_\text{int} (\tau_3), \zeta_k (\tau) \zeta_q (\tau) \zeta_p (\tau) \right]\right] \right] \right> \, ,
\end{eqnarray}
where $H_\text{int}$ is the interaction Hamiltonian given by \cite{Barnaby2011, Ferreira2014a}
\begin{eqnarray}
H_\text{int} = \frac{\xi}{2} \int d^3 x \, \zeta \,F_{\mu \nu} \tilde{F}^{\mu \nu}\, .
\end{eqnarray}
We are interested in the parametric dependence of the correlators with the temperature in order to compare it to previous results. For that reason we will ignore factors of $2\pi$ and other ${\cal O}(1)$ numbers.  We will assume the result to be dominated by the thermal part of the gauge field propagators. We also neglect the commutator structure, which should be an overestimation, since  there could be cancelations due to its structure. 

Let us first analyze the vertex structure. Generically, the 1-loop correction to the n-point function will be proportional to 
\begin{eqnarray}
\left< \left(F_{\mu \nu} \tilde{F}^{\mu \nu} \right)^n  \right>.
\end{eqnarray}
As we mentioned in the text, in the thermal regime we expect $\left< F \tilde{F} \right>$ to be suppressed because parity tends to be restored. In that case, the correlators involving the gauge fields will be dominated by $\left< FF \right>$ terms. By noting that $  \tilde{F}_{\mu \nu} \tilde{F}^{\mu \nu} = F_{\mu \nu} F^{\mu \nu}$ we estimate
\begin{eqnarray}
\left< \left(F_{\mu \nu} \tilde{F}^{\mu \nu} \right)^n  \right> \, \simeq \,
\begin{cases}
\left<F F\right>^n \qquad, \, \text{if n is even} \\
\left< F  \tilde{F} \right> \left<F F \right>^{n-1} \qquad, \, \text{if n is odd}
\end{cases}
\end{eqnarray}
up to permutation factors which we neglected here. The term in $\left< F F \right>$ is dimensionally an energy density and, using eq. (\ref{number}), we approximate in the thermal regime $\left<F F  \right> \simeq 4 k N(\omega_+(k))$, barring cancellations. On the other hand, the term in $\left< F \tilde{F} \right>$ is proportional to the parity asymmetry $\left< F_{\mu \nu} (k) \tilde{F}^{\mu \nu} (k) \right> = k d/d\tau (|A_+|^2-|A_-|^2) \simeq  d/d\tau ( N(\omega_+(k))- N(\omega_-(k) )$ where $\omega_{+,-}(k)=(k^2 \pm 2k\xi/ \tau)^{1/2}$. Assuming a thermal distribution we find ,in limit of large temperature, and for $-k \tau \gg \xi$
\begin{eqnarray}
\frac{d}{d \tau} \left[ N(\omega_+(k))- N(\omega_-(k)) \right] \simeq  \frac{4 \xi T}{H k^2 \tau^3} \, ,
\end{eqnarray}

Note, however, that in the in-in computation the vertices are evaluated at different times and so they could connect different physical momenta. Nevertheless, because the integrals are regulated in the UV by the particle number and in the IR by the momentum dependence of the integrand, we expect the integrals to peak when the gauge fields running in the loop have physical momenta of the order of the temperature $s_\text{phys} \simeq T$. 

Under these approximations and after Fourier transforming all the interaction Hamiltonians and integrating over the delta functions, the 1-loop correction simplifies to
\begin{eqnarray}
\left< \zeta_k \zeta_q \zeta_p \right>_\text{1-loop} (\tau) & \simeq & -\xi^3 \frac{\left(P_\zeta^\text{vac}\right)^3}{k^6} \, \delta^{(3)}\left(\vec{k}+\vec{q}+\vec{p}\right)  \int d^3 s \int_{-\infty}^{\tau} d \tau_1 \, \tau_1 s N(\omega_+(s))|_{\tau_1}  \int_{-\infty}^{\tau_1} d \tau_2 \, \tau_2 s N(\omega_+(s))|_{\tau_2}  \nn \\
&& \int_{-\infty}^{\tau_2} d \tau_3 \,\tau_3 \frac{d}{d \tau} \left[ N(\omega_+(s))- N(\omega_-(s)) \right]|_{\tau_3} \, ,
\end{eqnarray}
where we assumed the integrals to peak in the equilateral configuration, $k\simeq q \simeq p$, similarly to previous results in the non-thermal case \cite{Barnaby2011}, although there might be some changes in the thermal regime. In the last expression we have also replaced $\zeta$'s in the time integrals by their subhorizon expression $\zeta_k = H^2 \tau/ (\dot{ \phi} \sqrt{k})$, while the external ones by $\zeta_k = H^2/ (\dot{ \phi} \sqrt{k^3})$,  because they are evaluated at horizon crossing. Finally, to keep the treatment simple we further assume the integral to peak at a scale $s \gg k,q,p$. We could also include corrections with $s \simeq k,q,p$ but we expect not to have higher powers of the temperature so we neglect them for this estimation.

The time integrals should peak when the $s/a = -s H \tau = T$. At that scale the particle number is $N_{+,-}(T) \simeq 1$ and so we get
\begin{eqnarray} \label{NGint}
\left< \zeta_k \zeta_q \zeta_p \right>_\text{1-loop} (\tau) = - \frac{\xi^3 \left(P_\zeta^\text{vac}\right)^3}{k^6} \, \delta^{(3)}\left(\vec{k}+\vec{q}+\vec{p}\right) \int d^3 s \left(\frac{4 \xi }{s }\right)  \left[ \left(\frac{T}{2 s H}  \right)^2 s  \right]^2 \, ,
\end{eqnarray}
Therefore, the final result is parametrically given by  
\begin{eqnarray} \label{NGresult}
\left< \zeta_k \zeta_q \zeta_p \right>_\text{1-loop} (\tau) =  c \, \xi^4  \frac{\left(P_\zeta^\text{vac}\right)^3}{k^6}   {\cal O} \left( \frac{T^4}{ H^4}\right) \, \delta^{(3)}\left(\vec{k}+\vec{q}+\vec{p}\right)\, ,
\end{eqnarray}
where $c$ is a small number that contains inverse powers of $(2\pi)$ and we neglected a $\ln(T/H)$ term. This result means that the non-Gaussian parameter $f_{NL} \simeq \left< \zeta^3 \right>/\left< \zeta^2 \right>^2$ would be
\begin{eqnarray}
f_{NL} \simeq c \, \xi^4  P_\zeta^\text{vac}  {\cal O}\left( \frac{T^4}{H^4}\right)\, .
\end{eqnarray}
One should now compare this with the result in the non-thermal regime, where $f_{NL} \simeq 10^{-7} e^{6\pi \xi} P_\zeta^\text{vac}/\xi^8$ \cite{Barnaby2011}, keeping in mind that the constant $c$ could also be a small number. Let us look at a case of instantaneous thermalization at a temperature $\bar{T}$, where the energy in the plasma is the same as the energy before thermalization $\rho_{\g} \simeq 10^{-4} H^4 e^{2\pi \xi}/\xi^3$. Therefore, $\bar{T} \simeq \rho^{1/4} \simeq 0.1 H e^{\pi \xi/2}/\xi^{3/4}$. This means that, even at the very high temperature $\bar{T}$, non-Gaussianity is proportional to $e^{2\pi \xi}$ and so it is suppressed by $e^{-4\pi \xi}$ compared to the non-thermal case. Using the observational bound $f_{NL} < {\cal O}(10)$, we get $\xi \lesssim 6 - 7$ (assuming $10^{-7}\lesssim c \lesssim 10^{-4}$).
If we put instead the temperature $T_{eq}=\xi/\bar{g}$, derived in section~\ref{thermalmass}, due to the competition between the thermal mass and the instability, we get $\xi\lesssim 40-80$, using the same range of $c$ and using $\bar{g}=0.5$. 

In the case where inflaton perturbations are thermalized one should replace the $\zeta$ propagator by its thermal counterpart, which amounts to replace $1/2$ by $1/2+N_u$. A simple bound can be derived by noting that $N_u(k)\leq T/H$, for modes inside the horizon. Therefore, 
\begin{eqnarray}
\left< \zeta_k \zeta_q \zeta_p \right>^\text{thermal}_\text{1-loop} \lesssim d \, \xi^4  \frac{\left(P_\zeta^\text{vac}\right)^3}{k^6}   {\cal O} \left( \frac{T^4}{ H^4}\right) \, \delta^{(3)}\left(\vec{k}+\vec{q}+\vec{p}\right)\, \times \, \frac{T^3}{H^3}\, .
\end{eqnarray}
where $d$ is another small constant.
This implies that $f^{\text{thermal}}_{NL} =   \left< \zeta^3 \right>^\text{therm}/(\left< \zeta^2 \right>^\text{therm})^2$ would then be given by 
\begin{eqnarray} \label{NGresult2}
f^{\text{thermal}}_{NL}  \lesssim d \, \xi^4  P_\zeta^\text{vac}  {\cal O}\left( \frac{T^5}{H^5}\right)\, .
\end{eqnarray}
This result is only an upper bound, while a proper calculation could turn out to give something smaller.

Although this derivation is rough and a more proper study should be done, it shows parametrically that the loop corrections are strongly suppressed compared to the vacuum case. Apart from the technical caveats hidden in the approximations there is another caveat to this computation. As we argued in the text there are deviations for the thermal bath at low momenta and in particular we might expect a peak at those scales (in the case of an equilibrium temperature $T_{eq}$, the peak would be located at $\xi H$, as we argued in section~\ref{thermalmass}). In this computation we neglected the effect of such a peak on the non-Gaussianity. 

\subsection{Negligible superhorizon sourcing of curvature perturbation \label{iso}}

In this section we show that the superhorizon sourcing of curvature perturbation due to the presence of the gauge fields is negligible. The amount of superhorizon sourcing is given by
\begin{eqnarray}
\dot{ \zeta} (k)= - H \frac{\delta p_{\rm nad}}{\rho + p }
\end{eqnarray}
where $\delta p_{\rm nad} $ is the non-adiabatic pressure perturbation, $\rho$ is the total energy density and $p$ the total pressure. The non-adiabatic pressure in the case where the energy in the gauge fields is assumed to be subdominant is given by \cite{Nurmi:2013gpa}
\begin{eqnarray}
\delta p_{\rm nad}= \delta p_\gamma - \frac{\dot{p}}{\dot{ \rho}} \delta \rho_\gamma 
\end{eqnarray}
where $\gamma$ denotes the pressure and energy associated with the gauge fields. To leading order in slow-roll, and using $p_\gamma=1/3 \rho_\gamma$, then $\delta p_{\rm nad}=4/3 \rho_\gamma (k)$. On the other hand, 
\begin{eqnarray}
\rho+p = \frac{2}{3} \epsilon \rho_{\phi} + \frac{4}{3} \rho_\gamma.
\end{eqnarray}
Therefore, assuming  $ \rho_\gamma \ll  \epsilon \rho_{\phi} $ one gets
\begin{eqnarray}
\dot{\zeta} \simeq - 2 H \frac{\delta \rho_\gamma }{\epsilon \rho_\phi}.
\end{eqnarray}
At horizon crossing we know that the gauge fields can imprint a sizable effect in $\zeta$. Here, however, we are only interested in computing the sourcing on superhorizon scales. Therefore, we integrate the previous equation from a time $\tau_i$ where the modes are already superhorizon, $-k \tau_i <1$. The superhorizon correction to the 2-point of $\zeta$ in Fourier space is then given by
\begin{eqnarray}
\left< \zeta_k (\tau) \zeta_p (\tau) \right > = \left< \zeta_k (\tau) \zeta_p (\tau) \right >_* + 4 \int_{\tau_i}^\tau  \frac{d \tau_1 d \tau_2}{\tau_1 \tau_2} \frac{1}{\epsilon(\tau_1) \rho_\phi (\tau_1)} \frac{1}{\epsilon(\tau_2) \rho_\phi (\tau_2)} \left( \left< \delta \rho_\gamma (k, \tau_1) \delta  \rho_\gamma (p, \tau_2) \right> + \text{c.c.} \right)
\end{eqnarray}
where the star denotes the quantity evaluated at horizon crossing. For simplicity, and because that does not affect the leading order result, we assume $\epsilon$ and $\rho_\phi$ to be roughly constant. Then, the computation boils down to compute $\left< \delta  \rho_\gamma(k) \delta  \rho_\gamma(p) \right>$ where
\begin{eqnarray}
\delta \rho_\gamma (k,\tau)= \frac{1}{2(2\pi)^3 a^4} \int d^3 q \left[A'_q A'_{k-q} + q |k-q| A_q A_{k-q} \right] \vec{e}(\vec{q}) \cdot \vec{e}(\vec{k}-\vec{q}).
\end{eqnarray}
and we consider here only the $+$ polarization.
When computing the 2-point function of $\delta \rho_\gamma$ several permutations will appear but here we just want to show that the result is suppressed and so, for that reason, we just use the magnetic part of $\delta \rho_\gamma$. Moreover, both in the thermal and non-thermal regime, the bulk of the energy is in modes with momenta $-q\tau \gtrsim H$. Therefore, because we are evaluating the time integrals at times where $ -k \tau \ll 1 $ then, the integrals will peak at momenta $-q \tau \gg -k \tau$ and so $q-k \simeq q$. Therefore, after performing the contractions we are left with
\begin{eqnarray}
\left< \delta \rho_\gamma (k, \tau_1) \delta \rho_\gamma(p, \tau_2) \right> \approx \frac{ \tau_1^4 \tau_2 ^4 H^8}{2} \delta^{(3)}(\vec{p}+\vec{k}) \int d^3 q \, q^4 \left[ A_q (\tau_1)  A_q (\tau_2)^*  \right]^2. 
\end{eqnarray}
We now specify the computation to the non-thermal case where the analytical solutions are easier to handle. In the thermal case the reasoning is similar and would lead to a similar result.

In the non-thermal case the integrals peak at horizon crossing which means when $\tau_1 \simeq \tau_2 \simeq -1/q$. Therefore, by approximating the times to be the same we arrive at 
\begin{eqnarray}
\left< \zeta_k (\tau) \zeta_p (\tau) \right >  \approx  \left< \zeta_k (\tau) \zeta_p (\tau) \right >_* +  2 \frac{H^8}{\epsilon^2 \rho_\phi^2} \left(|A_q|^4 q^2 \right)_*  \int_{\tau_i}^\tau  d \tau_1 \, \tau^2_1    \delta^{(3)}(\vec{p}+\vec{k}). 
\end{eqnarray}
Therefore, while the first term is proportional to $k^{-3}$ to give the standard scale invariance, the second terms is proportional to $\tau^3$ where $-k\tau \ll 1$. Therefore we conclude that the superhorizon sourcing of curvature perturbation is negligible.

\bibliographystyle{JHEP}
\bibliography{thermalinflation}

\providecommand{\href}[2]{#2}\begingroup\raggedright\begin{thebibliography}{10}

\bibitem{Berera1995}
A.~Berera, {\it {Warm inflation}},  {\em Phys. Rev. Lett.} {\bf 75} (1995)
  3218--3221, [\href{http://arxiv.org/abs/astro-ph/9509049}{{\tt
  astro-ph/9509049}}].

\bibitem{Berera2008}
A.~Berera, I.~G. Moss, and R.~O. Ramos, {\it {Warm Inflation and its
  Microphysical Basis}},  {\em Rept. Prog. Phys.} {\bf 72} (2009) 026901,
  [\href{http://arxiv.org/abs/0808.1855}{{\tt arXiv:0808.1855}}].

\bibitem{Morikawa:1984dz}
M.~Morikawa and M.~Sasaki, {\it {Entropy Production in the Inflationary
  Universe}},  {\em Prog. Theor. Phys.} {\bf 72} (1984) 782.

\bibitem{Sakagami:1984ae}
M.-a. Sakagami and A.~Hosoya, {\it {Fate of Order Parameter in the Inflationary
  Universe}},  {\em Phys. Lett.} {\bf 150B} (1985) 342--346.

\bibitem{Anber2009}
M.~M. Anber and L.~Sorbo, {\it {Naturally inflating on steep potentials through
  electromagnetic dissipation}},  {\em Phys. Rev.} {\bf D81} (2010) 043534,
  [\href{http://arxiv.org/abs/0908.4089}{{\tt arXiv:0908.4089}}].

\bibitem{Barnaby2011}
N.~Barnaby, R.~Namba, and M.~Peloso, {\it {Phenomenology of a Pseudo-Scalar
  Inflaton: Naturally Large Nongaussianity}},  {\em JCAP} {\bf 1104} (2011)
  009, [\href{http://arxiv.org/abs/1102.4333}{{\tt arXiv:1102.4333}}].

\bibitem{Linde2012}
A.~Linde, S.~Mooij, and E.~Pajer, {\it {Gauge field production in supergravity
  inflation: Local non-Gaussianity and primordial black holes}},  {\em Phys.
  Rev.} {\bf D87} (2013), no.~10 103506,
  [\href{http://arxiv.org/abs/1212.1693}{{\tt arXiv:1212.1693}}].

\bibitem{Ferreira2014a}
R.~Z. Ferreira and M.~S. Sloth, {\it {Universal Constraints on Axions from
  Inflation}},  {\em JHEP} {\bf 12} (2014) 139,
  [\href{http://arxiv.org/abs/1409.5799}{{\tt arXiv:1409.5799}}].

\bibitem{Ferreira2015a}
R.~Z. Ferreira, J.~Ganc, J.~Noreña, and M.~S. Sloth, {\it {On the validity of
  the perturbative description of axions during inflation}},  {\em JCAP} {\bf
  1604} (2016), no.~04 039, [\href{http://arxiv.org/abs/1512.06116}{{\tt
  arXiv:1512.06116}}]. [Erratum: JCAP1610,no.10,E01(2016)].

\bibitem{Peloso2016}
M.~Peloso, L.~Sorbo, and C.~Unal, {\it {Rolling axions during inflation:
  perturbativity and signatures}},  {\em JCAP} {\bf 1609} (2016), no.~09 001,
  [\href{http://arxiv.org/abs/1606.00459}{{\tt arXiv:1606.00459}}].

\bibitem{Notari2016}
A.~Notari and K.~Tywoniuk, {\it {Dissipative Axial Inflation}},  {\em JCAP}
  {\bf 1612} (2016), no.~12 038, [\href{http://arxiv.org/abs/1608.06223}{{\tt
  arXiv:1608.06223}}].

\bibitem{Bartolo2014}
N.~Bartolo, S.~Matarrese, M.~Peloso, and M.~Shiraishi, {\it {Parity-violating
  CMB correlators with non-decaying statistical anisotropy}},  {\em JCAP} {\bf
  1507} (2015), no.~07 039, [\href{http://arxiv.org/abs/1505.02193}{{\tt
  arXiv:1505.02193}}].

\bibitem{Lin:2012gs}
C.-M. Lin and K.-W. Ng, {\it {Primordial Black Holes from Passive Density
  Fluctuations}},  {\em Phys. Lett.} {\bf B718} (2013) 1181--1185,
  [\href{http://arxiv.org/abs/1206.1685}{{\tt arXiv:1206.1685}}].

\bibitem{Bugaev:2013fya}
E.~Bugaev and P.~Klimai, {\it {Axion inflation with gauge field production and
  primordial black holes}},  {\em Phys. Rev.} {\bf D90} (2014), no.~10 103501,
  [\href{http://arxiv.org/abs/1312.7435}{{\tt arXiv:1312.7435}}].

\bibitem{Sorbo2011}
L.~Sorbo, {\it {Parity violation in the Cosmic Microwave Background from a
  pseudoscalar inflaton}},  {\em JCAP} {\bf 1106} (2011) 003,
  [\href{http://arxiv.org/abs/1101.1525}{{\tt arXiv:1101.1525}}].

\bibitem{Freese:1990rb}
K.~Freese, J.~A. Frieman, and A.~V. Olinto, {\it {Natural inflation with pseudo
  - Nambu-Goldstone bosons}},  {\em Phys. Rev. Lett.} {\bf 65} (1990)
  3233--3236.

\bibitem{Berera:2004vm}
A.~Berera, {\it {Warm inflation solution to the eta problem}},
  \href{http://arxiv.org/abs/hep-ph/0401139}{{\tt hep-ph/0401139}}.
  [PoSAHEP2003,069(2003)].

\bibitem{Cheng:2015oqa}
S.-L. Cheng, W.~Lee, and K.-W. Ng, {\it {Numerical study of pseudoscalar
  inflation with an axion-gauge field coupling}},  {\em Phys. Rev.} {\bf D93}
  (2016), no.~6 063510, [\href{http://arxiv.org/abs/1508.00251}{{\tt
  arXiv:1508.00251}}].

\bibitem{Adshead2015}
P.~Adshead, J.~T. Giblin, T.~R. Scully, and E.~I. Sfakianakis, {\it
  {Gauge-preheating and the end of axion inflation}},  {\em JCAP} {\bf 1512}
  (2015), no.~12 034, [\href{http://arxiv.org/abs/1502.06506}{{\tt
  arXiv:1502.06506}}].

\bibitem{Tkachev:1986tr}
I.~I. Tkachev, {\it {Coherent scalar field oscillations forming compact
  astrophysical objects}},  {\em Sov. Astron. Lett.} {\bf 12} (1986) 305--308.
  [Pisma Astron. Zh.12,726(1986)].

\bibitem{Anber2006}
M.~M. Anber and L.~Sorbo, {\it {N-flationary magnetic fields}},  {\em JCAP}
  {\bf 0610} (2006) 018, [\href{http://arxiv.org/abs/astro-ph/0606534}{{\tt
  astro-ph/0606534}}].

\bibitem{Schwinger:1951nm}
J.~S. Schwinger, {\it {On gauge invariance and vacuum polarization}},  {\em
  Phys. Rev.} {\bf 82} (1951) 664--679.

\bibitem{Hayashinaka:2016qqn}
T.~Hayashinaka, T.~Fujita, and J.~Yokoyama, {\it {Fermionic Schwinger effect
  and induced current in de Sitter space}},  {\em JCAP} {\bf 1607} (2016),
  no.~07 010, [\href{http://arxiv.org/abs/1603.04165}{{\tt arXiv:1603.04165}}].

\bibitem{Tangarife:2017rgl}
W.~Tangarife, K.~Tobioka, L.~Ubaldi, and T.~Volansky, {\it {Dynamics of Relaxed
  Inflation}},  \href{http://arxiv.org/abs/1706.03072}{{\tt arXiv:1706.03072}}.

\bibitem{Adshead:2016iae}
P.~Adshead, J.~T. Giblin, T.~R. Scully, and E.~I. Sfakianakis, {\it
  {Magnetogenesis from axion inflation}},  {\em JCAP} {\bf 1610} (2016) 039,
  [\href{http://arxiv.org/abs/1606.08474}{{\tt arXiv:1606.08474}}].

\bibitem{Brandenberger1993}
V.~F. Mukhanov, H.~A. Feldman, and R.~H. Brandenberger, {\it {Theory of
  cosmological perturbations. Part 1. Classical perturbations. Part 2. Quantum
  theory of perturbations. Part 3. Extensions}},  {\em Phys. Rept.} {\bf 215}
  (1992) 203--333.

\bibitem{Bartrum:2013fia}
S.~Bartrum, M.~Bastero-Gil, A.~Berera, R.~Cerezo, R.~O. Ramos, and J.~G. Rosa,
  {\it {The importance of being warm (during inflation)}},  {\em Phys. Lett.}
  {\bf B732} (2014) 116--121, [\href{http://arxiv.org/abs/1307.5868}{{\tt
  arXiv:1307.5868}}].

\bibitem{Linde2005}
A.~D. Linde, V.~Mukhanov, and M.~Sasaki, {\it {Post-inflationary behavior of
  adiabatic perturbations and tensor-to-scalar ratio}},  {\em JCAP} {\bf 0510}
  (2005) 002, [\href{http://arxiv.org/abs/astro-ph/0509015}{{\tt
  astro-ph/0509015}}].

\bibitem{future}
R.~Z. Ferreira and A.~Notari, {\it In preparation}, .

\bibitem{Ade:2015ava}
{\bf Planck} Collaboration, P.~A.~R. Ade {\em et~al.}, {\it {Planck 2015
  results. XVII. Constraints on primordial non-Gaussianity}},  {\em Astron.
  Astrophys.} {\bf 594} (2016) A17,
  [\href{http://arxiv.org/abs/1502.01592}{{\tt arXiv:1502.01592}}].

\bibitem{Mukhanov:2007zz}
V.~Mukhanov and S.~Winitzki, {\em {Introduction to quantum effects in
  gravity}}.
\newblock Cambridge University Press, 2007.

\bibitem{Nurmi:2013gpa}
S.~Nurmi and M.~S. Sloth, {\it {Constraints on Gauge Field Production during
  Inflation}},  {\em JCAP} {\bf 1407} (2014) 012,
  [\href{http://arxiv.org/abs/1312.4946}{{\tt arXiv:1312.4946}}].

\end{thebibliography}\endgroup

%%%%
\end{document}